# Relativistic Spacetime Crystals

Venkatraman Gopalan

Department of Materials Science and Engineering, Department of Physics, Department of Engineering Science and Mechanics, and the Materials Research Institute, The Pennsylvania State University, University Park, Pennsylvania 16802, USA.

Periodic space crystals are well established and widely used in physical sciences. Time crystals have been increasingly explored more recently, where time is disconnected from space. Periodic relativistic spacetime crystals on the other hand need to account for the mixing of space and time in special relativity through Lorentz transformation, and have been listed only in 2-dimensions. This work shows that there exists a transformation between the conventional Minkowski spacetime (MS) and what is referred to here as *renormalized blended spacetime* (RBS); they are shown to be equivalent descriptions of relativistic physics in flat spacetime. There are two elements to this reformulation of MS, namely, *blending* and *renormalization*. When observers in two inertial frames adopt each other's clocks as their own, while retaining their original space coordinates; the observers become *blended*. This process reformulates the Lorentz boosts into Euclidean rotations while retaining the original spacetime hyperbola describing worldlines of constant spacetime length from the origin. By renormalizing the blended coordinates with an appropriate factor that is a function of the relative velocities between the various frames, the hyperbola is transformed into a Euclidean circle. With these two steps, one obtains the RBS coordinates complete with new light lines, but now with a Euclidean construction. One can now enumerate the RBS point and space groups in various dimensions with their mapping to the well-known space crystal groups. The RBS point group for flat isotropic RBS spacetime is identified to be that of cylinders in various dimensions: **mm**2 which is that of a rectangle in 2D, (∞/**m**)**m** which is that of a cylinder in 3D, and that of hypercylinder in 4D. An antisymmetry operation is introduced that can swap between space-like and time-like directions, leading to color spacetime groups. The formalism reveals RBS symmetries that are not readily apparent in the conventional MS formulation. Mathematica® script is provided for plotting the MS and RBS geometries discussed in the work.

**1. Minkowski spacetime (MS), (*x, ct*)**

The goal of this work is to illustrate a transformation between the conventional flat relativistic spacetime (also called the Minkowski spacetime, MS, whose geometry is hyperbolic), and what is referred to here as *renormalized blended spacetime* (RBS, whose geometry is Euclidean). This will then form the basis for a mapping of the RBS crystals to the well-known space crystals, which in turn will help enumerate the former. To achieve this, we first briefly introduce the MS, followed by two critical steps required to reformulate it into RBS, namely, *blending* and *renormalization*. The former will largely retain the structure of the MS except to describe it with Euclidean angles and functions instead of hyperbolic angles and functions. The latter will transform the hyperbola into a circle. We largely adopt a geometric approach to special relativity and work in the early sections with 2-dimensional spacetime to keep the treatment accessible.

The geometry of a *Euclidean* 2-dimensional (2D) space spanned by unit vectors $\boldsymbol{x}$ and $\boldsymbol{y}$ possesses a norm (square) that is positive, i.e. $\boldsymbol{x} \cdot \boldsymbol{x} = \boldsymbol{y} \cdot \boldsymbol{y} = 1$. (Bold font is used for vectors and non-bolded font for coordinates). In two-dimensional space, the length, *r* of a vector, $\boldsymbol{r}$, from the origin to a point P is invariant under linear orthogonal transformations such as Euclidean rotations, inversion, or mirror. Given the coordinates (*x, y*) of the point P in the unprimed Euclidean coordinate system, and $(x', y')$ in the primed Euclidean coordinate system that shares the same origin and is related to the unprimed coordinate system by a linear orthogonal transformation, the length of the vector, $r$ will remain invariant, i.e.

$$x^2 + y^2 = x'^2 + y'^2 = r^2 \qquad (1)$$

In contrast, the geometry of special relativity is hyperbolic as described elegantly by Dray. [1] **Figure 1** schematically defines the three inertial frames of relevance in this work, which for pedagogical purposes, we label as the ground frame (GF), the train frame (TF) and the bird frame (BF). The TF and BF move at a velocity of $v$ and $u$ relative to the GF, respectively. Two inertial observers, one in the GF and another in the TF (depicted by the silhouette of girls depicted on the ground and on the moving train, respectively in **Fig. 1**) are observing an event (the bird flying) whose coordinates are measured in the GF as ($x$, $ct$), and in the TF as ($x'$, $ct'$), where $c$ is the speed of light in vacuum. The hyperbolic angles $\alpha$ and $\beta$ can be defined by the relative frame velocities, given by $v/c = tanh\alpha$ and $u/c = tanh\beta$. A geometric construction illustrating the significance of the hyperbolic angles is shown in **Figure 2**. The frame co-moving with the event (i.e. flying with the bird, or so called the bird frame, BF in **Fig. 1**) is typically called the *proper* frame, or the *wristwatch* frame.

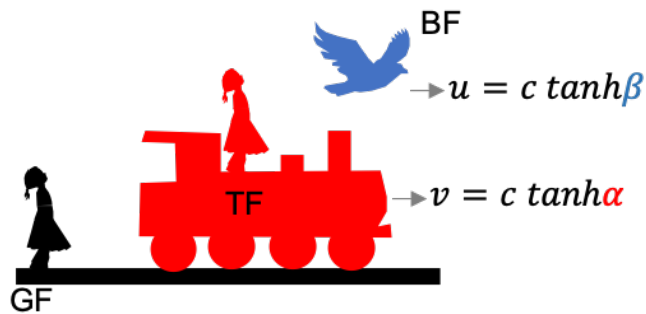

**Figure 1**. (a) The schematic depicts the stationary ground frame (GF, *x-ct*) observer. With respect to the GF, the train frame (TF, *x'-ct'*) observer moves with a velocity $v$ in the $+x$ direction. With respect to GF, an event (a bird) frame (BF) moves at a velocity $u$ in the $+x$ direction. The hyperbolic angles ($\alpha$ and $\beta$) are defined by the velocities, $u$ and $v$ relative to $c$ as indicated, and are illustrated in **Fig. 2**.

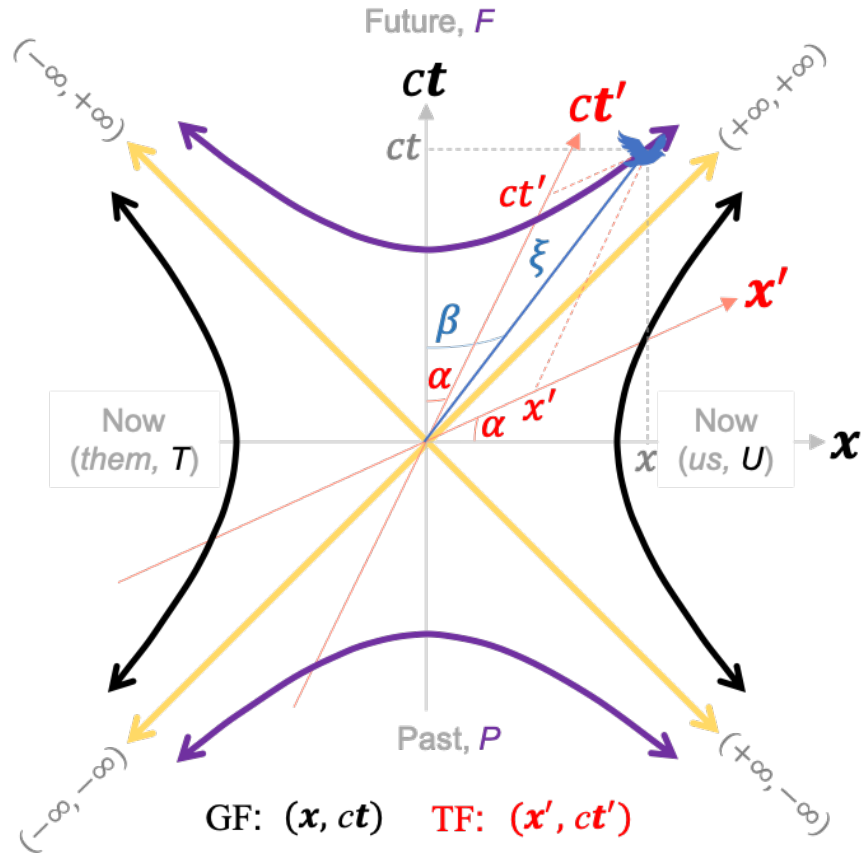

**Figure 2**. A 2D real Minkowski spacetime depicts hyperbolas given by $x^2 - (ct)^2 = x'^2 - (ct')^2 = \pm\xi^2$, where the purple pair of hyperbolas correspond to $-\xi^2$ (time-like directions from the origin) and the black pair of hyperbolas to $+\xi^2$ (space-like directions from the origin). An arbitrary time-like *event* is shown by a blue line from the origin to the event (the blue bird), and the projection of its coordinates $(x, ct) = (\xi \sinh\beta, \xi \cosh\beta)$ and $(x', ct') = (\xi \sinh(\beta -$

$\alpha$), $\xi cosh(\beta - \alpha)$) are depicted by broken lines on to the ground (GF, black) and the train (TF, red) frames. The diagonal yellow lines are the light lines given by $\xi = 0$; their poles, $+(\infty, \pm\infty)$ and $-(\infty, \pm\infty)$ are indicated. The four hyperbola branches are labeled *F*, *P*, *U* and *T*. See the *Mathematica®* script in the **Appendix A** or the *Mathematica®* notebook in the supplementary materials to generate this plot.

In 2D conventional relativistic spacetime spanned by unit vectors $\boldsymbol{x}$ (space axis) and $\boldsymbol{t}$ (time axis), $\boldsymbol{x} \cdot \boldsymbol{x} = -\boldsymbol{t} \cdot \boldsymbol{t} = 1$ (note the minus sign). In other words, if two inertial observers, GF and TF, moving at a relative velocity of $v$ to the GF (**Fig. 1**), observe the same event (bird) and record its coordinates as $(x, ct)$ and $(x', ct')$, respectively, then,

$$x^2 - (ct)^2 = x'^2 - (ct')^2 = \pm\xi^2, \qquad (2)$$

Where $\xi$ is called the spacetime *length*, $\xi^2$ is called the spacetime *interval*, $+\xi^2$ corresponds to spacetime directions from the origin along which *space-like events* occur (The *them, T* and *us, U*, hyperbola branches shown with black lines in **Fig. 2** represent such events with a constant spacetime length) and $-\xi^2$ to directions from the origin where the *time-like* events occur (the *future, F* and *past, P,* hyperbola branches shown as purple lines in **Fig. 2** represent such events with a constant spacetime length). Equation (2) thus describes hyperbola branches in the $\boldsymbol{x}$ - $\boldsymbol{ct}$ plane for a fixed $\xi$. In flat spacetime, $\xi^2$ is invariant across all inertial frames, i.e. independent of their relative velocity $v$. In 2D, Lorentz transformation relates the coordinates of an event (the bird) between a ground frame, $(x, t)$ and a train frame, $(x', t')$ moving along the $+\boldsymbol{x}$ axis with a speed of $v$, as follows:

$$\begin{pmatrix} ct' \\ x' \end{pmatrix} = \begin{pmatrix} \cosh\alpha & -\sinh\alpha \\ -\sinh\alpha & \cosh\alpha \end{pmatrix} \begin{pmatrix} ct \\ x \end{pmatrix}$$

$$= \begin{pmatrix} \gamma_v & -\gamma_v v/c \\ -\gamma_v v/c & \gamma_v \end{pmatrix} \begin{pmatrix} ct \\ x \end{pmatrix}$$

$$= \Lambda \begin{pmatrix} ct \\ x \end{pmatrix}, \tag{3}$$

In Eq. 3, $\cosh\alpha = \gamma_v = 1/\sqrt{1 - v^2/c^2}$, $\sinh\alpha = \gamma_v v/c$, and hence, $\tanh\alpha = v/c$. Furthermore, $\Lambda$, a 2×2 matrix with a determinant of 1, represents the Lorentz *boost*. It is also readily confirmed that Eqs. 2 and 3 are consistent.

In an effort to place space and time on an equal footing, Poincare' [2] and later Minkowski [3] defined an imaginary time ($ct \rightarrow ict$) such that a spacetime interval is defined now as $x^2 + (ict)^2$. Clearly, $x^2 + (ict)^2 = x'^2 + (ict')^2 = \pm\xi^2$ looks like a Euclidean norm and is identical to Eq. (2). However, Misner, Thorne and Wheeler bid "farewell to *ict*" in their classic book, *Gravitation*, [4] providing several reasons for doing so: suppression of the underlying metric structure ($(+\,-)$ in the 2D spacetime), hiding the distinction between covariant and contravariant quantities, hiding the interlocking causal structure imposed by the light cones, and not being generalizable to curved spacetime. Pedagogically, an imaginary time is somewhat non-intuitive.

Several authors in the past have proposed geometric constructions (see Guillaume [5], Mirimanoff [6], and Gruner [7,8] for its historical roots) that avoid imaginary time, and instead use real space and time coordinates. One such construction by Enrique Loedel Palumbo in 1948 [9] was rediscovered [10] independently by Henri Amar in 1955, and later re-rediscovered independently by Robert W. Brehme in 1961. [11] This construction (referred to here as LAB construction) makes the *choice* to draw the axes $x \perp ct'$ and $x' \perp ct$, a construction we will revisit next.

## 2. Blended spacetime coordinates, (*x, ct'*) and (*x', ct*) yield a Euclidean geometry

Rearranging terms in (3), one arrives at the following:

$$\begin{pmatrix} ct' \\ x \end{pmatrix} = \begin{pmatrix} sech\alpha & -tanh\alpha \\ tanh\alpha & sech\alpha \end{pmatrix} \begin{pmatrix} ct \\ x' \end{pmatrix}$$

$$= \begin{pmatrix} \frac{1}{\gamma_v} & -\frac{v}{c} \\ \frac{v}{c} & \frac{1}{\gamma_v} \end{pmatrix} \begin{pmatrix} ct \\ x' \end{pmatrix}$$

$$= R \begin{pmatrix} ct \\ x' \end{pmatrix} \qquad (4)$$

This represents a Lorentz transformation between $(x, t)$ and $(x', t')$ coordinates. Together, they are referred to here as a pair of *blended coordinates* composing a *blended spacetime*. These blended coordinates can be thought of as two inertial observers adopting each other's clock readings, while each retains their original inertial spatial coordinates. (Equivalently, they can adopt each other's spatial coordinates while retaining their own clocks). This can trivially be performed in a passive manner, post-measurement, assuming each observer knows special relativity and the two have an agreed upon origin. By redefining $\frac{1}{\gamma_v} = \cos\theta$, $\frac{v}{c} = \sin\theta$ and $\frac{\gamma_v v}{c} = \tan\theta$, we can rewrite Eq. (4) as follows:

$$\begin{pmatrix} ct' \\ x \end{pmatrix} = \begin{pmatrix} \cos\theta & -\sin\theta \\ \sin\theta & \cos\theta \end{pmatrix} \begin{pmatrix} ct \\ x' \end{pmatrix} = R \begin{pmatrix} ct \\ x' \end{pmatrix} \qquad (5)$$

Further, by rearranging Eq. (2), we get,

$$x^2 + (ct')^2 = x'^2 + (ct)^2 = \eta^2 \tag{6}$$

If we define $ds'^{\circ 2} = (ct)^2 + x'^2$ and the $ds^{\circ'2} = (ct')^2 + x^2$ as the spacetime intervals in the blended coordinates, we gather from Eq. (6) that $ds'^{\circ 2} = ds^{\circ'2}$. These intervals describe the Euclidean interval between the event and the origin in the blended spacetime frames $(x, ct')$ and $(x', ct)$, generated by the blending of GF and the TF observers in **Figure 1**. This looks like a Euclidean measure. The Euclidean interval $\eta^2$ is however *not* an invariant across different inertial frames in the MS; it is a function of both $v$ and $u$, as derived next.

If we write $\eta = \xi\chi$, then Equation (6) motivates us to define *blended Euclidean* coordinates as follows:

$$(x, ct') = \xi\chi(sin\phi, cos\phi),$$
$$(x', ct) = \xi\chi(sin(\phi - \theta), cos(\phi - \theta)) \tag{7a}$$

Here, the angle definitions are: $\frac{u}{c} = sin\phi/cos(\phi - \theta)$ (for events along time-like directions in MS), $\frac{u}{c} = cos(\phi - \theta)/sin\phi$ (for events along space-like directions in MS), and $\frac{v}{c} = sin\theta$. In other words,

$$\frac{u}{c} = min\left(\frac{x}{ct}, \frac{ct}{x}\right) \tag{7b}$$

Note in particular that these definitions ensure that $v, u \leq c$.

To find an expression for $\chi$ as a function of the Euclidean angles, we substitute the coordinates of Eq. (7) into Eq. (2) for events observed from the GF, namely, $x^2 - (ct)^2 = \pm\xi^2$. One finds that $\xi^2\chi^2(sin^2\phi - cos^2(\phi - \theta)) = \pm\xi^2$; here the positive sign is for space-like directions and the negative sign for time-like directions. Upon simplification, this leads to $\chi^2 = \mp sec\theta \, sec(2\phi - \theta)$, where the negative sign is for space-like directions and the positive sign for time-like directions. Alternately, one could substitute the hyperbolic coordinates of a general event from Fig. 2 into equation (6) to show that $\chi^2 = cosh\alpha \, cosh(2\beta - \alpha) > 0$, since a *cosh* function is always positive. One could therefore equivalently write $\chi^2 = |sec\theta \, sec(2\phi - \theta)|$, (in order to ensure that it stays positive for all Euclidean angles) and hence,

$$\chi = +\sqrt{|sec\theta \, sec(2\phi - \theta)|}$$
$$= \sqrt{cosh\alpha \, cosh(2\beta - \alpha)}$$
$$= +\gamma_v\gamma_u/\sqrt{|1 - \gamma_u u(\gamma_u u - 2\gamma_v v)/c^2|} \qquad (8)$$

Here the positive root is chosen without a loss of generality, and $\gamma_u = 1/\sqrt{1 - u^2/c^2}$.

In a similar fashion, substituting Eq. (7) into Eq. (2) for events observed from the TF, $x'^2 - (ct')^2 = \pm\xi^2$, we get the same expression for $\chi$ as noted above. The term $\chi$ is called the *renormalization factor*, and is plotted in **Figure 3** as a function of $u/c$ for three different values of $v/c$, namely, $v = 0$, $v = 0.9c$, and $v = u$. These three cases will be explored further in the following sections. The light lines are the vertical asymptotes at $u/c = \pm 1$ where the $\chi$ diverges (i.e. $\chi \to \infty$).

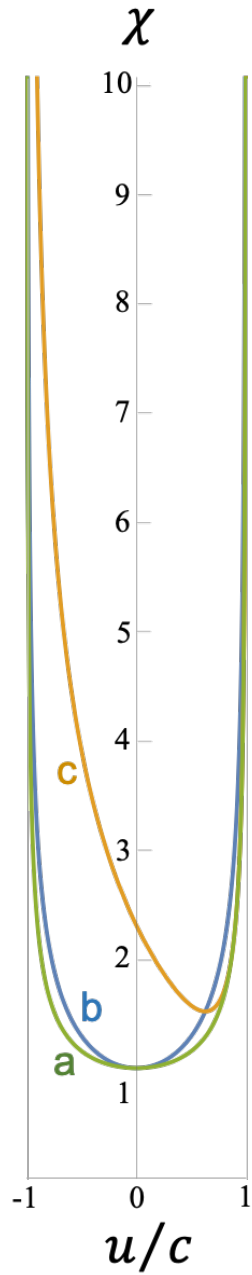

**Figure 3**: Plots of the renormalization factor $\chi$ From Eq. (8) as a function of $u/c = sin\phi/cos(\phi - \theta)$ for (a) $v = u$, $\phi = \theta$ (b) $v = 0$, $\theta = 0$, and (c) $v = 0.9c$. The light lines correspond to the vertical asymptotes at $u/c = \pm 1$. See the *Mathematica®* script in **Appendix A** or the *Mathematica®* notebook in the supplementary materials to generate this plot.

With the Euclidean coordinates in Eqs. (7) plus (8) in hand, we are ready to replot the MS in **Fig. 2** in terms of the blended and the RBS coordinates. **Figure 4** plots the coordinates of Eq. (7) (along with Eq. (8)) for the special case of $\theta = 0$. This is the case of a stationary train in **Fig. 1**, with $v = 0$. Strikingly, one can capture all the four hyperbolas in **Fig. 2** including the time-like and space-like events by varying $\phi$ (bird flying at varying speeds, $u$). When $\theta = 0°$, the plot reproduces the hyperbolas and the light lines shown in **Fig. 2** with the $x$ and $x'$ coordinates coincident (horizontal axis), $ct$ and $ct'$ coincident (vertical axis), and $x \perp ct$. This mathematical exercise is important since it shows that the hyperbolas in the MS can be captured equally well with Euclidean functions and angles in **Fig. 4**, instead of hyperbolic functions and angles as in **Fig. 2**.

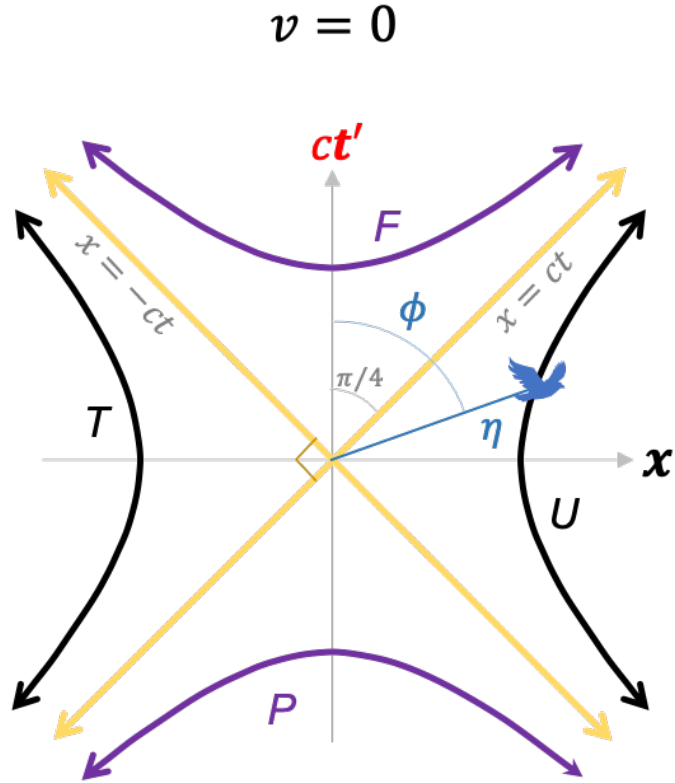

**Figure 4**: A plot of the 2D Euclidean blended spacetime coordinates in Eqs. (7) with Eq. (8) substituted in it, for $v = c \sin\theta = 0$. The hyperbolas in **Fig. 2** are recovered but the angles are now Euclidean. For this case, the $x$ and $x'$ coordinates are coincident (horizontal axis), and the $ct$ and $ct'$ axes are coincident (vertical axis), and $x \perp t$. See the *Mathematica®* script in **Appendix A** or the *Mathematica®* notebook in the supplementary materials to generate this plot.

However, when $v \neq 0$ as shown in **Figure 5**, the hyperbolas are rotated by a Euclidean rotation angle $\theta$ which captures the Lorentz boost, Eq. (5), between the two pairs of blended coordinates. The light lines given by $x = \pm ct$ results in the condition, $sin\phi = \pm cos(\phi - \theta)$, which, for example for $v/c = sin\theta = 0.9$, yields the orientations of the light lines as $\phi = 77.079°$ and $\phi = -12.921°$ as shown in **Fig. 5**.

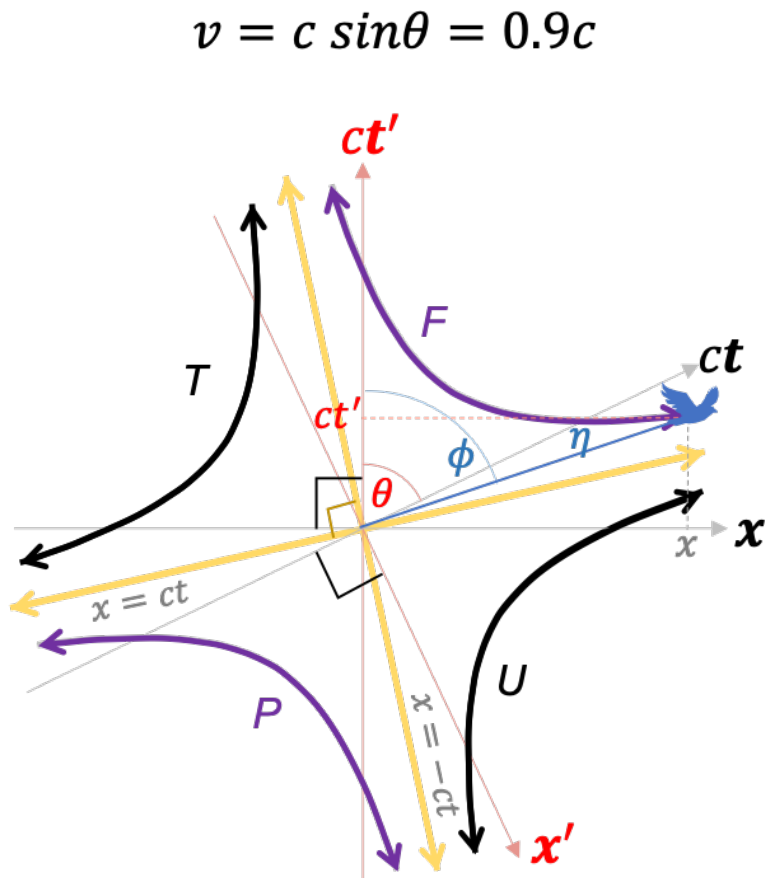

$$v = c\ sin\theta = 0.9c$$

**Figure 5**: A plot of the 2D Euclidean blended spacetime coordinates from Eqs. (7) and (8) for $v = c\,sin\theta = 0.9c$. The two light lines are oriented at the angles of $\phi = 77.079°$ and $\phi = -12.921°$. See the *Mathematica®* script in **Appendix A** or the *Mathematica®* notebook in the supplementary materials to generate this plot.

## 3. Renormalized Blended spacetime (RBS) coordinates

Rearranging Eq. (7), it is clear that

$$\frac{1}{\chi}(x, ct') = \frac{\xi\chi}{\chi}(sin\phi, cos\phi),$$

$$\frac{1}{\chi}(x', ct) = \frac{\xi\chi}{\chi}(sin(\phi - \theta), cos(\phi - \theta)) \quad (9a)$$

Note that we are intentionally not "canceling out" the $\chi$ terms on the right-hand side of Eq. (9), since $\chi \to \infty$ when $u \to c$. In that special case, we should consider the limit as follows:

$$\lim_{u \to c, \chi \to \infty} \frac{\xi\chi}{\chi}(sin\phi, cos\phi) = \xi(sin\phi, cos\phi) \text{ and,}$$

$$\lim_{u \to c, \chi \to \infty} \frac{\xi\chi}{\chi}(sin(\phi - \theta), cos(\phi - \theta)) = \xi(sin(\phi - \theta), cos(\phi - \theta)) \quad (9b)$$

If we define the renormalized coordinates as follows:

$$\bar{x} = x/\chi, \quad \bar{t} = t/\chi,$$

$$\bar{x}' = x'/\chi, \quad \bar{t}' = t'/\chi, \quad (10)$$

Then, the *RBS coordinates* can be rewritten as:

$$(\bar{x}, c\bar{t}') = \frac{\xi\chi}{\chi}(sin\phi, cos\phi),$$

$$(\bar{x}', c\bar{t}) = \frac{\xi\chi}{\chi}(sin(\phi - \theta), cos(\phi - \theta)) \tag{11a}$$

Again, in the limit of $\chi \to \infty$ when $u \to c$, one has to take the limits on the right-hand side using the L'Hôpital's rule, $\lim_{\chi \to \infty} \frac{\chi}{\chi} = 1$, leading to the following:

$$(\bar{x}, c\bar{t}') = \xi(sin\phi, cos\phi),$$

$$(\bar{x}', c\bar{t}) = \xi(sin(\phi - \theta), cos(\phi - \theta)) \tag{11b}$$

The Lorentz transformation in Eq. (4) can now be rewritten in the RBS coordinates as:

$$\begin{pmatrix} c\bar{t}' \\ \bar{x} \end{pmatrix} = \begin{pmatrix} cos\theta & -sin\theta \\ sin\theta & cos\theta \end{pmatrix} \begin{pmatrix} c\bar{t} \\ \bar{x}' \end{pmatrix} \tag{11c}$$

Equation (6) can be rewritten as an *RBS invariant* as:

$$\bar{x}^2 + (c\bar{t}')^2 = \bar{x}'^2 + (c\bar{t})^2 = \xi^2 \tag{12}$$

Where we take the limit $\lim_{u \to c, \chi \to \infty} \frac{\xi^2 \chi^2}{\chi^2} = \xi^2$ on the right-hand side. This provides the equation of a circle in the RBS coordinates. This construction is equivalent to the LAB construction [9–11] where the *choice* made to draw the axes $\boldsymbol{x} \perp \boldsymbol{ct'}$ and $\boldsymbol{x'} \perp \boldsymbol{ct}$ is implicit in the Euclidean coordinate choice in Eq. (7). Consider next, four special cases of the RBS coordinates, namely $v = 0$, $v = 0.9c$, $v \to c$ and $v = u$.

***Case I, $v = 0$ ($\theta = 0$)***: Here, the GF and the TF observers are coincident; this could be considered as the limit where the GF observer is *self-blending*. Upon renormalization by $\chi$ according to Eqs. (10), the four hyperbola branches depicted in **Figure 4** transform into four arc segments of a circle as shown in **Fig. 6**, two of them time-like (purple segments, where $sec\theta\, sec(2\phi - \theta) > 0$), and the other two, space-like (black segments, where $sec\theta\, sec(2\phi - \theta) < 0$). This is essentially the case of a *renormalized Minkowski spacetime*, or RMS. Blending is essentially missing here; hence it is one of the simplest cases of "Euclideanizing" MS.

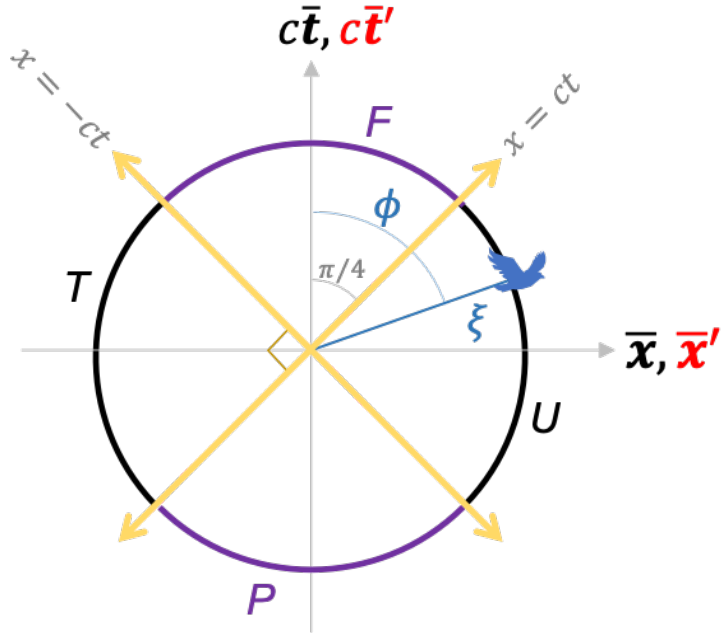

**Figure 6**: A special case of the blended and the RBS coordinates, Eq. (11), where $v = 0$. RBS coordinates, given in Eq. (11) transforms the blended coordinates plot in **Figure 4** into a circle of radius $\xi$, where the four hyperbola branches in **Figure 4** become four arc segments of the circle here. Purple (black) arc segments represent time-like (space-like) events. Light lines are shown by yellow lines. See the *Mathematica®* script in **Appendix A** or the *Mathematica®* notebook in the supplementary materials to generate this plot.

The RBS coordinates also possess RBS light lines as $u \to \pm c$. To see this, consider that the light lines are defined in the MS by $x = \pm ct$. When $u \to \pm c$, $\chi \to \infty$ from **Fig. 3**. From Eq. (11), $(\bar{x}, c\bar{t}) = (\xi \sin\phi, \xi \cos(\phi - \theta))$; hence the light lines correspond to the condition, $\sin\phi =$ $\pm cos(\phi - \theta)$. This equality has a solution for $\phi$ given any value of $\theta$. For example, when $v = c\, sin\theta = 0$, the two RBS light lines are at angles of $\phi = \pm \pi/4$ as shown in **Figure 6**. The corresponding coordinates for the light lines in the MS are therefore $(x, ct) = \chi(\xi sin\phi, \pm \xi sin\phi) \rightarrow +(\infty, \pm \infty)$ or $-(\infty, \pm \infty)$, which is consistent with the four infinity limits of the light lines in the hyperbolic construction in **Fig. 2**. Conversely, starting from the $(x, ct) = \chi(\xi sin\phi, \xi cos(\phi - \theta))$ coordinates in the MS and renormalizing with $\chi$ as shown in Eq. (9), one encounters a $\chi/\chi \rightarrow \infty/\infty$ term as $u \rightarrow \pm c$; However, there is a well-defined *limit* of $\lim_{\chi \rightarrow \infty}(\chi/\chi) = 1$; in this limit the RBS coordinates are $(\bar{x}, c\bar{t}) \rightarrow (\xi sin\phi, \xi cos(\phi - \theta))$. Furthermore, as $u \rightarrow \pm c$, $sin\phi = \pm cos(\phi - \theta)$; hence $(\bar{x}, c\bar{t}) \rightarrow \xi(1, \pm 1)$, which are the light lines shown in **Figure 6**. Thus, the light lines in the RBS coordinates (**Figure 6**) map to the $(\infty, \pm \infty)$ or $-(\infty, \pm \infty)$ limits of the light lines in the MS coordinates (**Figure 2**). We will more formally discuss these mappings in the next section.

A remarkable consequence of formulating this problem with the Euclidean angle $\phi$ is that it can be continuously varied from 0 to $2\pi$ around a circle *without* violating any relativistic physics. This means that one can smoothly "rotate across" the RBS light lines in **Fig. 6** that is not possible with the hyperbolic angle, $\beta$ in **Fig. 2**. This is because in the span that $\phi$ varies from 0-to-$\pi/4$, $\beta$ varies from 0-to-$\infty$, both of which correspond to approaching the light line. Note that $\phi = \pi/4$ results in a well-defined limit of $(\bar{x}, c\bar{t}) = (\xi sin\phi, \xi cos\phi) = (\xi/\sqrt{2})(1, 1)$; this point lies on the light line in **Fig. 6** just as expected, the same limit that was obtained earlier when $\beta \rightarrow \infty$ in **Fig. 2**. Now consider what happens when $\phi$ changes by an infinitesimal amount, $\epsilon$, from a value of $\frac{\pi}{4}$, which is a deviation from the RBS light line in either direction, i.e. $\phi = \frac{\pi}{4} \pm \epsilon$. Now, $(\bar{x}, c\bar{t}) = \frac{\xi}{\sqrt{2}}(cos\epsilon \pm sin\epsilon, cos\epsilon \mp sin\epsilon)$. As $\epsilon \rightarrow 0$ in a continuous manner, $(\bar{x}, c\bar{t}) \rightarrow \frac{\xi}{\sqrt{2}}(1,1)$, namely one mathematically approaches the light line smoothly as expected. Thus, the mathematical crossing across the RBS light line by varying $\phi$ is smooth and continuous. This is a big departure from the hyperbolic construction of spacetime in **Fig. 2,** where one is unable to mathematically "cross" the MS light lines by boosting an event frame, and hence has to "stay put" in one of the four hyperbolic branches for a finite spacetime length, $\xi$. We will have more to say about the formal mapping between the MS and RBS spaces in the next section.

*Case II, $v = 0.9c$:* In this case, the hyperbola branches in the blended coordinates in **Figure 5** transform into arcs of a circle in the RBS coordinates of Eq. (11). This is shown in **Figure 7**. The orientation of the RBS light lines is found by exploring the limit of $u \to \pm c$, $\chi \to \infty$ (see **Fig. 3**). From Eq. (11), $(\bar{x}, c\bar{t}) = (\xi sin\phi, \xi cos(\phi - \theta))$; hence the RBS light lines correspond to $sin\phi = \pm cos(\phi - \theta)$. When $v = c\, sin\theta = 0.9c$, the two RBS light lines are at angles of $\phi = 77.079°$ and $\phi = -12.921°$, respectively as shown in **Figure 7**. Interestingly, the RBS light lines rotate in the Euclidean plane as $v$ varies. This is explored further next.

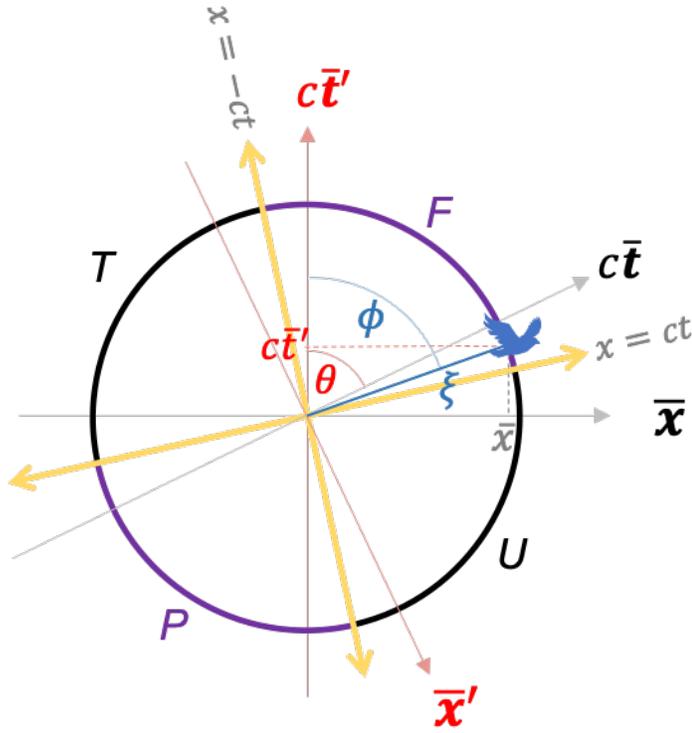

**Figure 7**: Renormalized Blended Spacetime (RBS) coordinates that turns the four hyperbolas (*F, P, U,* and *T*) in **Figure 5** to arcs of a circle. An arbitrary event (a bird) and its RBS coordinates are depicted. See the *Mathematica®* script in **Appendix A** or the *Mathematica®* notebook in the supplementary materials to generate this plot.

***Case III, $v \to c$:*** As $v \to c$, the angle $\theta \to \pi/2$. This is a case of blending between the GF and the TF where the latter is moving at $v \to c$. The resulting blended and RBS frame plots are shown in **Figure 8**. The light lines for this case can be found by setting $x = \pm ct$ and $x' = \pm ct'$. From the coordinates in Eq. (7) and in the limit of $\theta \to \pi/2$, one can therefore rewrite these relations as, $\xi\chi sin\phi = \pm \xi\chi sin\phi$ and $\xi\chi cos\phi = \mp\xi\chi cos\phi$. These relations imply that the RBS light lines correspond to $\phi \to \pm 0$ and $\phi \to \pm\frac{\pi}{2}$ as shown. As in the previous case, one can show that for $\phi = \frac{\pi}{2} \pm \epsilon$ or $\phi = \pm \epsilon$, the RBS coordinates smoothly approach the RBS light lines as $\epsilon \to 0$.

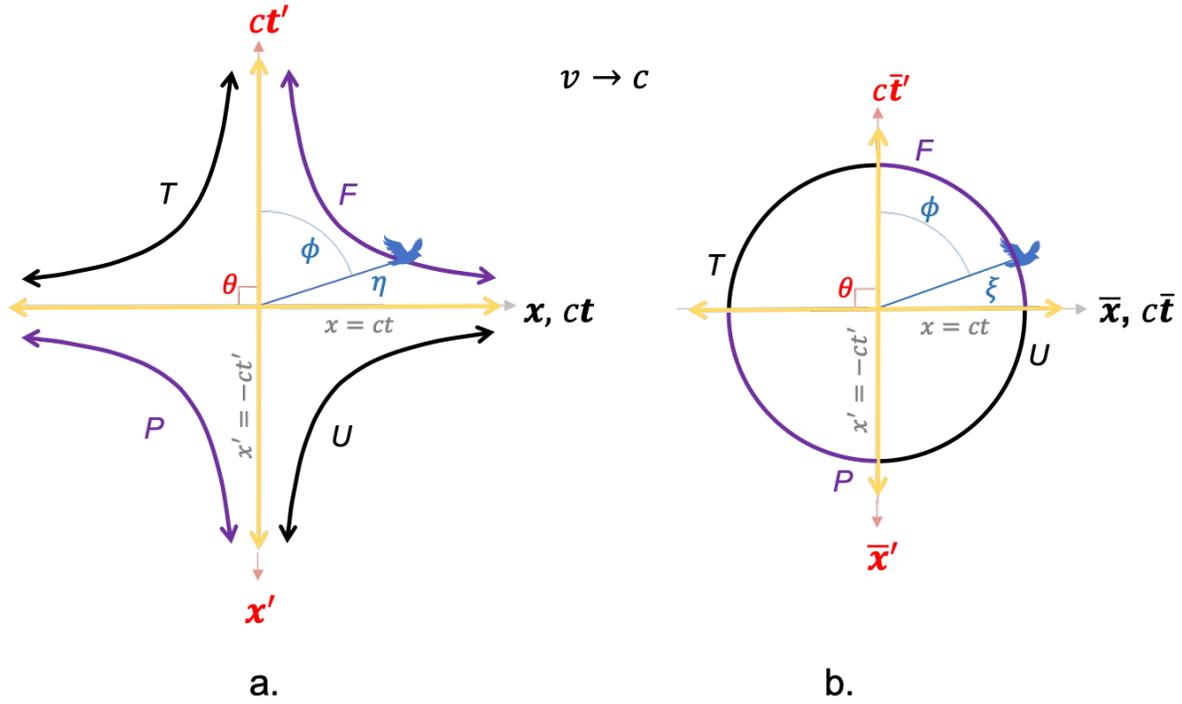

**Figure 8**: A special case of the blended and the RBS coordinates, where $v \to c$. (a) A plot of the blended coordinates given in Eqs. (7) and (8). (b) RBS coordinates, given in Eq. (11) transforms (a) into a circle of radius $\xi$, where the four hyperbola branches in (a) become four arc segments of the circle. Purple (black) hyperbolas branches and arc segments represent directions from the origin where events occur for a fixed $\xi$ in time-like (space-like) directions. Blended and RBS Light lines are shown by yellow lines. See the *Mathematica®* script in the **Appendix A** or the *Mathematica®* notebook in the supplementary materials to generate theses plots.

***Case IV, $v = u$:*** Here the TF and BF merge into each other, i.e. the case of a *proper* frame. This can also be deduced by noting that when $v = u$ in **Figure 5**, $\theta = \phi$, and the coordinate $\bar{x}' = 0$, which corresponds to the set of events on the $\bar{t}'$ axis in **Figure 2**; by definition, those events are occurring in the *proper* frame.

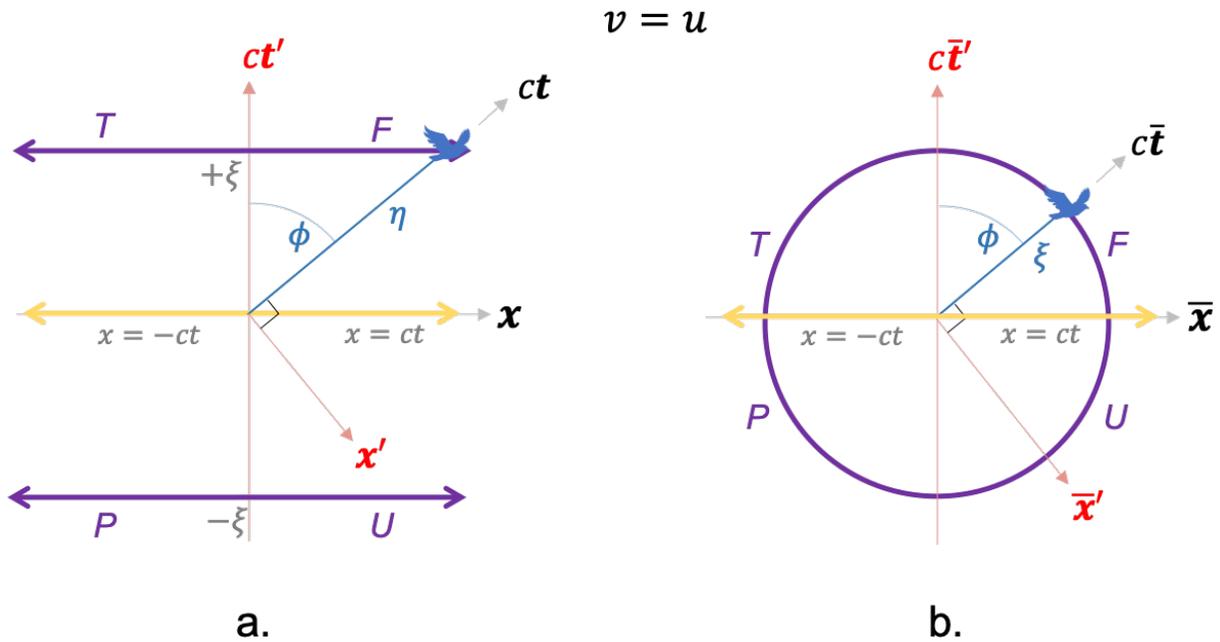

**Figure 9**: A special case of the blended and the RBS coordinates, where $v = u$, and hence $\theta = \phi$. (a) A plot of the blended coordinates given in Eq. (7) with Eq. (8) substituted in. (b) RBS coordinates, given in Eq. (11) transforms (a) into a circle of radius $\xi$, where the four hyperbola branches in (a) become four arc segments of the circle. Remarkably, all the arc segments now represent directions from the origin along which time-like events occur. Blended and RBS Light lines are shown by the horizontal yellow line. See the *Mathematica®* script in the **Appendix A** or the *Mathematica®* notebook in the supplementary materials to generate this plot.

When the GF and the BF are blended without renormalization, one gets the blended spacetime plot in **Figure 9a**. While in the other cases (I-III) discussed in the text, $v$ (and hence TF) could have been thought of as fixed while $u$ varied, in the case of $v = u$, the TF is moving along with the event. It is an unusual (but a mathematically allowed) case of a coordinate system $(x', ct)$ that is moving with the event frame in MS. In other words, let's say the GF girl observes an event 1 with a spacetime length of $\xi$ in the MS frame. This event becomes the "bird". Now she blends her coordinates with the proper coordinates in the BF of event 1. If she now observes a different event 2 with a spacetime length of $\xi$ but a different boost than event 1; she again repeats the process by blending with the proper frame of event 2. The GF is thus directly blending with the proper frame of *any* event she observes at a spacetime length of $\xi$ from the origin and with varying boosts.

In this special case, the hyperbola branches in the conventional spacetime in **Fig. 2** flatten into straight horizontal lines at $\pm\xi$. This is understood mathematically as follows: The plot of $(x, ct') = \xi\chi(sin\phi, cos\phi)$, where $\chi = +\sqrt{|sec\theta\ sec(2\phi - \theta)|}$ can be simplified for this case of $\phi = \theta$ to $\chi = |sec\phi|$. Hence, $(x, ct') = \xi|sec\phi|(sin\phi, cos\phi)$. The reason for the "flattening" of the $(x, ct')$ plots is due to the $|sec\phi|$ function, which diverges (i.e. $\to \infty$) at $\phi = \pm\pi/2$. Thus, the coordinate $x \to \infty$ diverges, while $ct' = \xi|sec\phi|cos\phi \to \pm 1$. This defines the two purple horizontal lines shown in Figure 9a.

When renormalized by $\chi$ according to Eq. (11), one gets a circle of radius $+\xi$ as shown in **Figure 9b**. Remarkably, all the events in both **Figs. 9a** and **b** are along time-like directions! This is seen by starting with the blended coordinates in Eq. (7) when $\theta = \phi$, namely, $(x, ct') = \xi\chi(sin\phi, cos\phi)$ and $(x', ct) = \xi\chi(0, 1)$, where from Eq. (8), $\chi = |sec\phi|$. By substituting in Eq. (2), one gets, $x^2 - (ct)^2 = x'^2 - (ct')^2 = -\xi^2\chi^2 cos^2\phi < 0$, which indicates a time-like direction.

Another unusual aspect of this case is that the two light lines merge into a single blended or RBS light line parallel to the $x$ axis, as shown. The light lines are defined by $x = \pm ct$, which is equivalent to $\xi\chi \sin\phi = \pm\xi\chi$, which suggests that $\phi \to \pm\pi/2$ as shown. The light lines are also defined by $x' = \pm ct'$, which implies $\xi\chi 0 = \pm\xi\chi\cos\phi$, which again yields $\phi \to \pm\pi/2$. This implies that the blended and RBS light lines coincide with the horizontal $x$-axis in Fig. 9a. This is perhaps the simplest and somewhat surprising RBS geometry one could imagine: a time-like circle of constant RBS interval with a single light line.

## 4. Mapping of events from the Minkowski to the RBS coordinates

Now we formally explore the transformation and the type of mapping between the MS and the RBS coordinates. We explore two cases: from MS→RBS (in this section), and from RBS→MS (in the next section). This will be used to validate that the RBS coordinates do indeed capture the relativistic physics content of the MS coordinates.

Consider the transformation from the conventional rest frame in the Minkowski spacetime (MS) to the renormalized blended spacetime (RBS) frames as follows:

$$\begin{pmatrix} c\bar{t} \\ \bar{x}' \end{pmatrix} = \frac{1}{\chi}\begin{pmatrix} 1 & 0 \\ -\gamma_v \frac{v}{c} & \gamma_v \end{pmatrix}\begin{pmatrix} ct \\ x \end{pmatrix} = \frac{1}{\chi}\Lambda'^{\circ}\begin{pmatrix} ct \\ x \end{pmatrix}, \qquad \text{and}$$

$$\begin{pmatrix} c\bar{t}' \\ \bar{x} \end{pmatrix} = \frac{1}{\chi}\begin{pmatrix} \gamma_v & -\gamma_v \frac{v}{c} \\ 0 & 1 \end{pmatrix}\begin{pmatrix} ct \\ x \end{pmatrix} = \frac{1}{\chi}\Lambda^{\circ\prime}\begin{pmatrix} ct \\ x \end{pmatrix} \qquad (13)$$

Similarly, the transformation from the moving frame, TF to the RBS frames is as follows:

$$\begin{pmatrix} c\bar{t} \\ \bar{x}' \end{pmatrix} = \frac{1}{\chi}\begin{pmatrix} \gamma_v & \gamma_v \frac{v}{c} \\ 0 & 1 \end{pmatrix}\begin{pmatrix} ct' \\ x' \end{pmatrix} = \frac{1}{\chi}\Lambda'^{\circ}\Lambda\begin{pmatrix} ct' \\ x' \end{pmatrix}, \qquad \text{and}$$

$$\begin{pmatrix} c\bar{t}' \\ \bar{x} \end{pmatrix} = \frac{1}{\chi}\begin{pmatrix} 1 & 0 \\ \gamma_v \frac{v}{c} & \gamma_v \end{pmatrix}\begin{pmatrix} ct \\ x \end{pmatrix} = \frac{1}{\chi}\Lambda^{\circ\prime}\Lambda\begin{pmatrix} ct' \\ x' \end{pmatrix} \tag{14}$$

Consider now starting from a general coordinate $(x, ct) = (\sigma, \lambda)$ in the conventional spacetime frames in **Fig. 2**. How do they transform into the blended coordinates? From Eq. (3), $(x', ct') = \left(-\lambda\gamma_v \frac{v}{c} + \sigma\gamma_v, \lambda\gamma_v - \sigma\gamma_v \frac{v}{c}\right)$. By renormalization with $\chi$, we get the blended coordinates, $(\bar{x}, c\bar{t}') = \frac{1}{\chi}\left(\sigma, \lambda\gamma_v - \sigma\gamma_v \frac{v}{c}\right)$ and $(\bar{x}', c\bar{t}) = \frac{1}{\chi}\left(-\lambda\gamma_v \frac{v}{c} + \sigma\gamma_v, \lambda\right)$. For $v \neq c$, every MS event thus has unique and well-defined RBS coordinates.

How about the events along the light lines, $x = \pm ct$? In this case, $\lambda = \pm\sigma$ for which $u = c$. Then, $(x', ct') = \sigma\gamma_v\left(1 \mp \frac{v}{c}, \pm\left(1 \mp \frac{v}{c}\right)\right)$, $(\bar{x}, c\bar{t}') = \frac{\sigma}{\chi}\left(1, \pm\gamma_v\left(1 \mp \frac{v}{c}\right)\right)$ and $(\bar{x}', c\bar{t}) = \frac{\sigma}{\chi}\left(\gamma_v\left(1 \mp \frac{v}{c}\right), \pm 1\right)$. From **Fig. 3**, as $u \to c$, $\chi \to \infty$. If $v \neq c$ and $\sigma, \lambda$ are finite, then $(\bar{x}, c\bar{t}') \to (0,0)$ and $(\bar{x}', c\bar{t}) \to (0,0)$. Thus, the events with finite coordinates on the two MS light lines in **Fig. 2** map to the RBS origin (such as in **Figs. 6** and **7**), a many-to one mapping from the MS to the RBS. This is summarized in **Figure 10**.

In linear algebra, this is expressed as follows: the *kernel* (or *nullspace*) of the transformation $\frac{1}{\chi}\Lambda^{\prime\circ}$ (and the transformation $\frac{1}{\chi}\Lambda^{\circ\prime}$) from the MS to the RBS coordinates is the set of all events that form the light-line in the MS, namely, the lines $x = \pm ct$. The *range* of the transformation matrix $\frac{1}{\chi}\Lambda^{\prime\circ}$ is the 2D blended vector space spanned by the column vectors of this transformation matrix, namely, $(0, \gamma_v/\chi) \equiv (0,1)$ and $\left(1/\chi, -\frac{\gamma_v}{\chi}\frac{v}{c}\right) \equiv \left(1, -\gamma_v \frac{v}{c}\right)$. The *domain* of the transformation is spanned by the column vectors of the inverse of the $\frac{1}{\chi}\Lambda^{\prime\circ}$ matrix.

What about the $(x, ct) = (\sigma, \lambda) \to \pm(\infty, \pm\infty)$ corresponding to the infinity limits of the light lines in the MS frame? This again corresponds to $u = c$, and hence $\chi \to \infty$. In the next section, it is shown that in the limit of $u = c$, the four infinity limits of the light lines, $(x, ct) \to +(\infty, \pm\infty)$ and $-(\infty, \pm\infty)$, map to finite, well-defined coordinates in the RBS. These results are also summarized in **Figure 10**.

## 5. Mapping of events from the RBS to the MS coordinates

Consider a general event coordinate given by $(\bar{x}, c\bar{t}') = (\delta, \Delta)$ in **Figure 6** or **7** in the RBS frame. Using Eqs. (4) and (10), one can determine the corresponding coordinates in the $(\bar{x}', c\bar{t})$ frame and in the MS frames as follows. From the definition of the normalized coordinates, it follows that $(x, ct') = \chi(\delta, \Delta)$. From Eq. (4), it follows that $(x', ct) = \chi\left(-\frac{\Delta v}{c} + \frac{\delta}{\gamma_v}, \frac{\Delta}{\gamma_v} + \frac{\delta v}{c}\right)$. Renormalizing for a finite $\chi$ according to Eq. (10), one can find that $(\bar{x}', c\bar{t}) = \frac{\chi}{\chi}\left(-\frac{\Delta v}{c} + \frac{\delta}{\gamma_v}, \frac{\Delta}{\gamma_v} + \frac{\delta v}{c}\right)$. For $v \neq c$, all of these coordinates are well defined, and there is a well-defined mapping from the RBS to the MS coordinates and between the two RBS frames.

How about the events along the light lines, $x = \pm ct$ in the RBS coordinates in **Figs. 6** or **7**? In this case, from above, $\chi\delta = \pm\chi\left(\frac{\Delta}{\gamma_v} + \frac{\delta v}{c}\right)$. Rearranging we get, $\Delta = \pm\frac{\chi}{\chi}\gamma_v\delta\left(1 \mp \frac{v}{c}\right)$. Substituting this relation into the MS coordinates above, we get, $(x, ct) = \chi\delta\left(1, \pm\frac{\chi}{\chi}\left(1 \mp \frac{v}{c}\right) + \frac{v}{c}\right)$. However, light lines correspond to $u = c$, and hence $\chi \to \infty$. Hence, $(x, ct) = \lim_{\chi \to \infty} \chi\delta\left(1, \pm\frac{\chi}{\chi}\left(1 \mp \frac{v}{c}\right) + \frac{v}{c}\right) = \chi\delta(1, \pm 1) \to (\infty, \pm\infty)$ or $-(\infty, \pm\infty)$, depending on the sign of $\delta$.

Thus, any arbitrary event $(\bar{x}, c\bar{t}') = \delta\left(1, \pm\gamma_v\left(1 \mp \frac{v}{c}\right)\right)$ on the light lines in the RBS frame (**Figs.**

6 or **7**) map to one of the four infinity limits, +(∞, ±∞) or -(∞, ±∞), of the light lines in the MS frame (**Fig. 2**) as shown in Section 4.

## 6. Summary of important results thus far leading to the RBS coordinates

We pause to summarize the relationships between the MS, blended spacetime, and the RBS. This is done through **Figure 10** where the important equations and representative diagrams are presented for each spacetime. The information content in all three frames in terms of relativistic physics is equivalent, i.e. all essential physics is captured in translating between these frames. For $v \neq c$, every MS event has unique and well-defined RBS coordinates. Light lines in MS frame map to the origin in the RBS frame, while the light lines in the RBS frame map to the $+(\infty, \pm\infty)$ and the $-(\infty, \pm\infty)$ poles in the MS frame. This is an example of a transformation, where points at infinity in MS are transformed to finite Euclidean points in the RBS. Both frames have a pair of equivalent light lines that capture the same physics. Among significant qualitative differences, MS does not allow for a mathematical "crossing" of the light line through a hyperbolic Lorentz boost, while this is possible in the RBS as shown above. This can be succinctly understood as follows: in hyperbolic geometry, a "time-like event" (i.e. event along a time-like direction) is represented by the coordinates $(x, ct) = \xi(sinh\beta, cosh\beta)$ which approaches $+(\infty, \pm\infty)$ and the $-(\infty, \pm\infty)$ as the event frame is boosted and it approaches the light lines; hence its coordinates diverge, and the event frame can only approach the light lines asymptotically. Using the RBS transformation, these infinity limits of the MS light lines can be transformed to finite Euclidean RBS coordinates, given by $(\bar{x}, c\bar{t}') = \delta\left(1, \pm\gamma_v\left(1 \mp \frac{v}{c}\right)\right)$. Now the RBS light lines can be "approached" and even "crossed" upon boosting an event frame.

There is no contradiction in the relativistic physics between the two formalisms. For example, consider the simple case of $v = u$ in the RBS coordinates as shown in **Figure 9**. This case leads to the condition, $\theta = \phi \pm m\pi$, where $m$ is an integer, and $u/c = \pm \sin\phi$, which places no restriction on the angle $\phi$. The RBS light lines in this case are at $\phi = \pm \pi/2$. When $\phi$ increases from zero to $\phi = \pi/2$, the event frame velocity, $u$ increases from zero to $u = c$. Upon crossing the RBS light line at $\phi = \pi/2$, when $\phi$ exceeds $\pi/2$, the $u$, according to equation 7(b), slows down back from $c$ and approaches a value of zero when $\phi = \pi$. Between $\pi \leq \phi \leq 3\pi/2$, the $u$ speeds up again to equal $-c$ upon approaching the light line at $\phi = 3\pi/2$. Finally, after "crossing" the RBS light line a second time, the event frame slows down again to zero upon reaching $\phi = 2\pi$. All of this is consistent with Einstein's postulates in flat spacetime; at no point does the speed, $u$, of the event frame exceed $c$.

Now consider the case when $v \neq u$. In all these cases in Figs. 4, 5, and 6 for example, there are two light lines. Consider the specific case of $v = 0.9c$ in Fig. 5 and recall that $v/c = \sin\theta$, and in accordance with Eq. 7(b), $u/c = \sin\phi / \cos(\phi - \theta)$ (for events along time-like directions), and $u/c = \cos(\phi - \theta)/ \sin\phi$ (for events along space-like directions). Starting from $\phi = 0$, which corresponds to $u = 0$, and traveling along the $F$ branch, upon reaching the first RBS light line at $\phi = 77.079°$, $u = c$. At this stage, the event switches from being in time-like directions to space-like directions on branch $U$, and $u$ starts decreasing back down from $c$. When $\phi = 90°$, $u = v$. Upon further travel along the $U$ branch, when $\phi = 90° + \theta$, $u = 0$. Continuing further on the $U$ branch and reaching the second RBS light line crossing at $\phi = 90° + 77.079°$, $u = -c$. As $\phi$ increases further, $u$ decreases, while the events are now along time-like directions again on the arc $P$. At $\phi = 180°$, $u = 0$. On further travel along the arc $P$, the speed $u$ increases again until it reaches $u = c$ at $\phi = 257.079°$ where it meets the RBS light line again. Beyond that, the events again switch to lying along space-like directions on the arc T. At $\phi = 270°$, $u = v$, and at $\phi = 270° + \theta$, $u = 0$. At $\phi = 347.079°$, we mathematically cross the RBS light line again, and $u = -c$. Beyond that, the events are again back on arc F along time-line directions. At $\phi = 360°$, $u = 0$, and we are back a full circle.

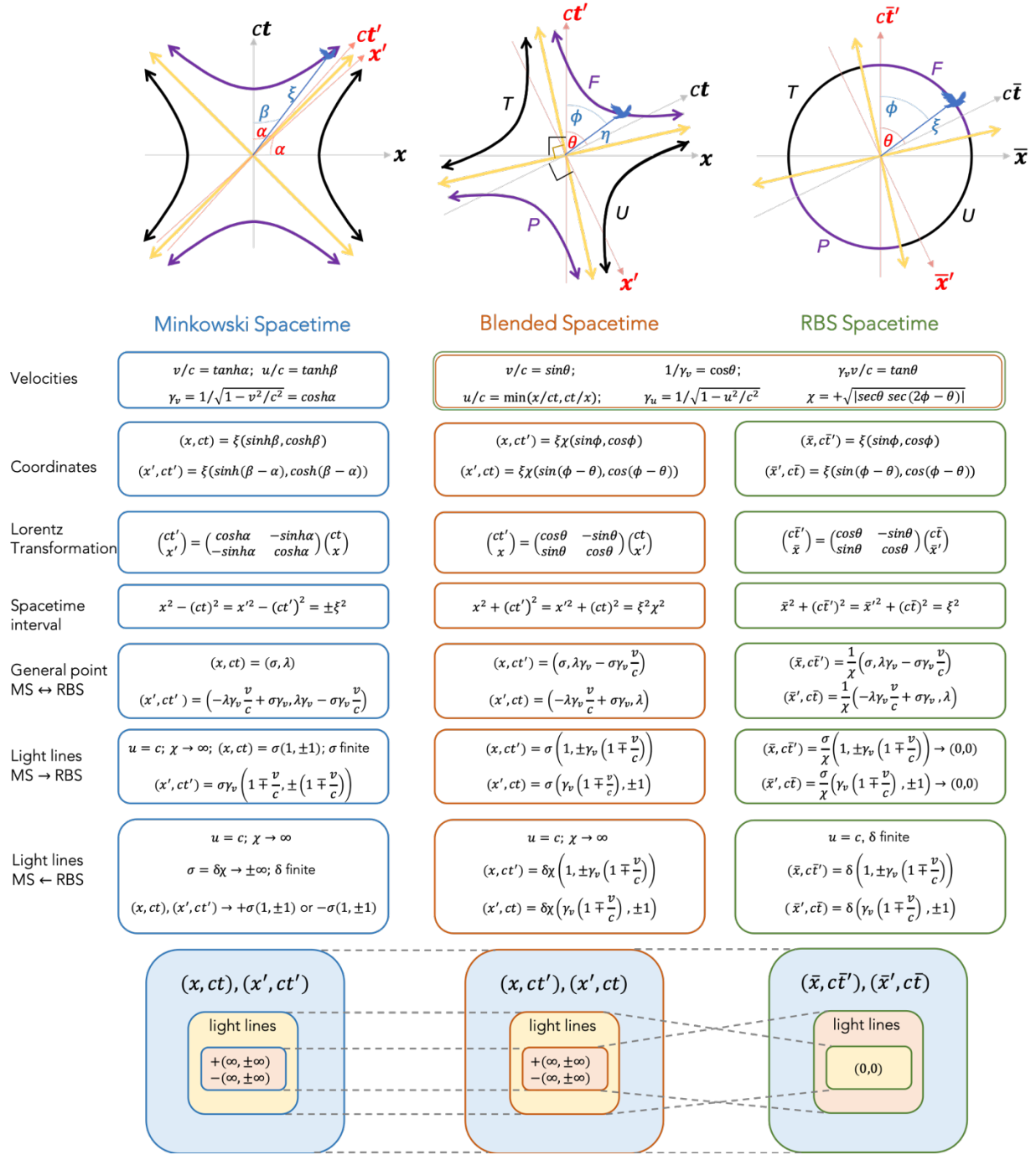

**Figure 10**: A summary of representative diagrams (top row diagrams simplified from **Figs. 2, 5, and 7** for $v = 0.9c$) and the key equations below mapping the Minkowski spacetime (MS), blended spacetime, and the renormalized blended spacetime (RBS). The bottom row schematics indicates the mappings between the three spacetimes shown by the dashed grey lines. In addition

to the MS, blended spacetime, and RBS coordinates, one could also express three more coordinates: Minkowski polar, $(\xi, \beta)$, blended polar, $(\eta, \phi)$, and renormalized blended polar, $(\xi, \phi)$. Relations between all six of these coordinate systems can be deduced from the above information and that given in the main text.

### 7. Lorentz and Poincaré' groups in the RBS coordinates

Consider now a generalization to the Minkowski 4D spacetime, defined by the three contravariant space coordinates, $x^1, x^2, x^3$, and time coordinate, $x^0 = ct$. The *proper Lorentz group*, $\mathcal{L}$, comprises of six operations within an isotropic 4D spacetime [12,13]: three independent Euclidean rotations, $(\theta^{12}, \theta^{23}, \theta^{31})$, respectively within one of the three space planes, $x^1 - x^2, x^2 - x^2, x^3 - x^1$. Their coordinate transformation matrices are respectively given by:

$$\begin{bmatrix} 1 & 0 & 0 & 0 \\ 0 & cos\theta^{12} & -sin\theta^{12} & 0 \\ 0 & sin\theta^{12} & cos\theta^{12} & 0 \\ 0 & 0 & 0 & 1 \end{bmatrix}, \begin{bmatrix} 1 & 0 & 0 & 0 \\ 0 & 1 & 0 & 0 \\ 0 & 0 & cos\theta^{23} & -sin\theta^{23} \\ 0 & 0 & sin\theta^{23} & cos\theta^{23} \end{bmatrix}, \begin{bmatrix} 1 & 0 & 0 & 0 \\ 0 & cos\theta^{31} & 0 & -sin\theta^{31} \\ 0 & 0 & 1 & 0 \\ 0 & sin\theta^{31} & 0 & cos\theta^{31} \end{bmatrix} \quad (15)$$

In addition, $\mathcal{L}$ has three independent Lorentz boosts, $(\alpha^{01}, \alpha^{02}, \alpha^{03})$ (similar to $\Lambda$, in Eq. (3)), each respectively within one of the space-time planes, $x^1 - ct, x^2 - ct, x^3 - ct$. Their coordinate transformation matrices are respectively given by:

$$\begin{bmatrix} cosh\alpha^{01} & -sinh\alpha^{01} & 0 & 0 \\ -sinh\alpha^{01} & cosh\alpha^{01} & 0 & 0 \\ 0 & 0 & 1 & 0 \\ 0 & 0 & 0 & 1 \end{bmatrix}, \begin{bmatrix} cosh\alpha^{02} & 0 & -sinh\alpha^{02} & 0 \\ 0 & 1 & 0 & 0 \\ -sinh\alpha^{02} & 0 & cosh\alpha^{02} & 0 \\ 0 & 0 & 0 & 1 \end{bmatrix}, \begin{bmatrix} cosh\alpha^{03} & 0 & 0 & -sinh\alpha^{03} \\ 0 & 1 & 0 & 0 \\ 0 & 0 & 1 & 0 \\ -sinh\alpha^{03} & 0 & 0 & cosh\alpha^{03} \end{bmatrix} \quad (16)$$

In the RBS frame, the *RBS proper Lorentz group*, $\mathcal{L}_{RBS}$, the same spatial rotations matrices as in Eq. (15) are valid, except in the $\bar{x}^1 - \bar{x}^2, \bar{x}^2 - \bar{x}^3, \bar{x}^3 - \bar{x}^1$ planes, respectively. However, one notices from Eq. (5), that the Lorentz boosts given in Eq. (16) can instead be written as Euclidean rotations. The three Lorentz boosts in Eq. (16) are now rewritten in the RBS frames as Euclidean rotations:

$$\begin{bmatrix} cos\theta^{01} & -sin\theta^{01} & 0 & 0 \\ sin\theta^{01} & cos\theta^{01} & 0 & 0 \\ 0 & 0 & 1 & 0 \\ 0 & 0 & 0 & 1 \end{bmatrix}, \begin{bmatrix} cos\theta^{02} & 0 & -sin\theta^{02} & 0 \\ 0 & 1 & 0 & 0 \\ sin\theta^{02} & 0 & cos\theta^{02} & 0 \\ 0 & 0 & 0 & 1 \end{bmatrix}, \begin{bmatrix} cos\theta^{03} & 0 & 0 & -sin\theta^{03} \\ 0 & 1 & 0 & 0 \\ 0 & 0 & 1 & 0 \\ sin\theta^{03} & 0 & 0 & cos\theta^{03} \end{bmatrix} \quad (17)$$

Here, $\theta^{0i}$ ($i \equiv 1,2,3$) are the three Euclidean Lorentz boost angles in the $\bar{x}^i - c\bar{t}'$ planes given by $sin\theta^i = v^i/c$. Typically, one defines these rotations in the range of $-c < v^i < c$, which translates to $-\pi/2 < \theta^i < \pi/2$. However, one is allowed to vary $\theta^i$ smoothly across the light lines in the RBS coordinates from $0 < \theta^i < 2\pi$, without violating any relativistic physics; the maximum $v^i$ will still remain $c$.

If now *RBS spacetime inversion*, $\bar{1}'_{RBS}$, is defined as $\bar{1}'_{RBS}: (c\bar{t}', \bar{x}^1, \bar{x}^2, \bar{x}^3) \to (-c\bar{t}', -\bar{x}^1, -\bar{x}^2, -\bar{x}^3)$, *RBS time reversal* as $1'_{RBS}: (c\bar{t}', \bar{x}^1, \bar{x}^2, \bar{x}^3) \to (-c\bar{t}', \bar{x}^1, \bar{x}^2, \bar{x}^3)$ and *RBS spatial inversion* as $\bar{1}_{RBS}(c\bar{t}', \bar{x}^1, \bar{x}^2, \bar{x}^3) \to (c\bar{t}', -\bar{x}^1, -\bar{x}^2, -\bar{x}^3)$, then, $\{1, \bar{1}'_{RBS}, 1'_{RBS}, \bar{1}_{RBS}\}$ forms a group, $\mathcal{I}$, where 1 stands for the identity matrix. Through a direct product of the *proper RBS Lorentz group* with the group $\{1, \bar{1}'_{RBS}, 1'_{RBS}, \bar{1}_{RBS}\}$, i.e., $\mathcal{L}_{RBS} \otimes \mathcal{I}$, a new group is created, called the *extended RBS Lorentz group*, $\mathcal{L}_{eRBS}$. [12,13] (A note on notation: in crystallography, $1'$ denotes time reversal; the superscript "prime" has nothing to do with the "prime" used to represent the train frame, TF, here in special relativity. Similarly, the overbar such as $\bar{1}$ in conventional crystallography is used to denote spatial inversion; it has nothing to do with the overbar used here for renormalization as in Eq. (10). The coincidence is unfortunate, but the context should clarify: the use of prime and overbar in conjunction with symmetry elements represent time reversal and spatial inversion, respectively. If the prime and overbar are used in conjunction with spacetime coordinates, as in Eq. (10), they represent TF and renormalization respectively.)

How about translations? So far, we have described spacetime intervals observed from a common origin by GF and TF observers in MS or by the blended observers in RBS. A general translation would move the origin, which would in general, rescale the spacetime interval for a given event. In an infinite space crystal with translational symmetry, there is a periodic placement of atoms. Space groups describe their global symmetry while point groups describe the local (site) symmetry at individual locations within the crystal. Similarly, one could create translational symmetry in a spacetime crystal with periodic placement of events, where the global symmetry is described by the *Poincare' space groups* and the local (site) symmetry is described by the Lorentz point groups. In such a case, a single observer (conventional or blended) at a selected origin would observe all of these infinite series of periodic events in the manner described so far with Lorentz groups. Just as in space crystals, translations would also create a periodic set of observers (conventional or blended) related by translational symmetry, each observing identical environment of events around them. The translational symmetry of spacetime captured by Poincare' groups is discussed next.

The *proper Poincare' group*, $\mathcal{P}$, in 4D MS coordinates consists of the proper Lorentz group, $\mathcal{L}$ combined with four translations, namely, $(x^\mu) + (T^0, 0,0,0)$, $(x^\mu) + (0, T^1, 0,0)$, $(x^\mu) + (0,0, T^2, 0)$ and $(x^\mu) + (0,0,0, T^3)$, where $(x^\mu) \equiv (x^0, x^1, x^2, x^3)$, and $T^\mu$ are the translations along the coordinates $\mu$ that vary from 0 to 3. [12,13] If in addition, improper transformations are included, namely *spatial inversion*, $(ct, x^1, x^2, x^3) \rightarrow (ct, -x^1, -x^2, -x^3)$, and *time reversal*, $(ct, x^1, x^2, x^3) \rightarrow (-ct, x^1, x^2, x^3)$, then one forms *extended Poincare' group*, $\mathcal{P}_e$ [12,13].

The *proper RBS Poincare' group*, $\mathcal{P}_{RBS}$, in 4D coordinates is similarly defined as the *proper RBS Lorentz group*, $\mathcal{L}_{RBS}$ plus four translations, namely, $(x_{RBS}^\mu) + (\bar{T}^{0'}, 0,0,0)$, $(x_{RBS}^\mu) + (0, \bar{T}^1, 0,0)$, $(x_{RBS}^\mu) + (0,0, \bar{T}^2, 0)$ and $(x_{RBS}^\mu) + (0,0,0, \bar{T}^3)$, where $(x_{RBS}^\mu) \equiv (\bar{x}^{0'}, \bar{x}^1, \bar{x}^2, \bar{x}^3)$, and $\bar{T}_{RBS}^\mu \equiv (\bar{T}^{0'}, \bar{T}^1, \bar{T}^2, \bar{T}^3)$ are the translations along the RBS coordinates, $(c\bar{t}', \bar{x}^1, \bar{x}^2, \bar{x}^3)$, respectively. If these translations are included in the *extended RBS Lorentz group*, $\mathcal{L}_{eRBS}$, one gets *extended RBS Poincare' group*. (A note again on duplicate notation: the word *proper* here is used again in a crystallographic sense as excluding mirrors and inversion. It has nothing to do with the *proper* used earlier in the context of the event frame.)

## 8. Two-dimensional RBS point groups

A striking mathematical consequence of this formulation is that the RBS *Lorentz and Poincare groups* can now be mapped to the Euclidean point and space groups for space crystals, respectively; the latter are all fully listed [14–16]. Space crystals in various dimensions can be classified into point and space groups: 17 space and 10 point groups in 2D; 230 space and 32 point groups in 3D; 4895 space and 271 point groups in 4D, and so on [14–16]. In contrast, to the best of my knowledge, only a handful of relativistic crystal groups (in 2D) have been listed so far [17].

Let us first begin with 2D *point* groups, so called because all the symmetry elements of the group must leave the coordinates of at least one point in the object or spacetime unchanged (invariant). In the discussion below, we will work from the RBS plots in **Figures 6, 7, 8b,** and **9b** in order to identify the relevant symmetry groups. We notice in these figures two features that are important to consider: Light lines, and events at a fixed RBS spacetime length of $\xi$ along space- versus time-like directions from the origin, represented by black and purple arc segments respectively. We consider black versus purple line segments to be related by a *color symmetry* as discussed further later. Consider the following cases:

*Colorless symmetry including all the features of the RBS diagrams:* If the RBS light lines and the distinction between space-versus time-like directions is paid attention to, one notices a point group symmetry of **mm2** in the RBS diagrams of **Figures 6, 7, 8b,** and **9b**. (Group labels are shown in bold font, while the elements of the group are shown not bolded; the term "colorless" recognizes the presence of black versus purple arc segments, but does not introduce a new symmetry element to switch between the two, as is done later on). This is depicted in **Figure 11**. The complete point group for **Figures 6, 7, 8b, 9b** is given as **mm2** $\equiv \{1, 2, m_{L1L2}, m_{\overline{L1}L2}\}$. The element 1 represents identity. The element 2 represents a 2-fold rotation (i.e. a rotation of $2\pi/2$) in the $(x, ct')$ plane. Note that such a proper 2-fold rotation transformation does not exist in the original MS construction; it is a hidden symmetry in the RBS. The two mirrors $m_{L1L2}$ and $m_{\overline{L1}L2}$ bisect the four quadrants formed by the RBS light lines, labeled by the subscripts L1 and L2 here. (Note that there is only one RBS light line in **Figure 9b**, hence one of the mirrors is parallel to the light lines, and the other perpendicular to it.)

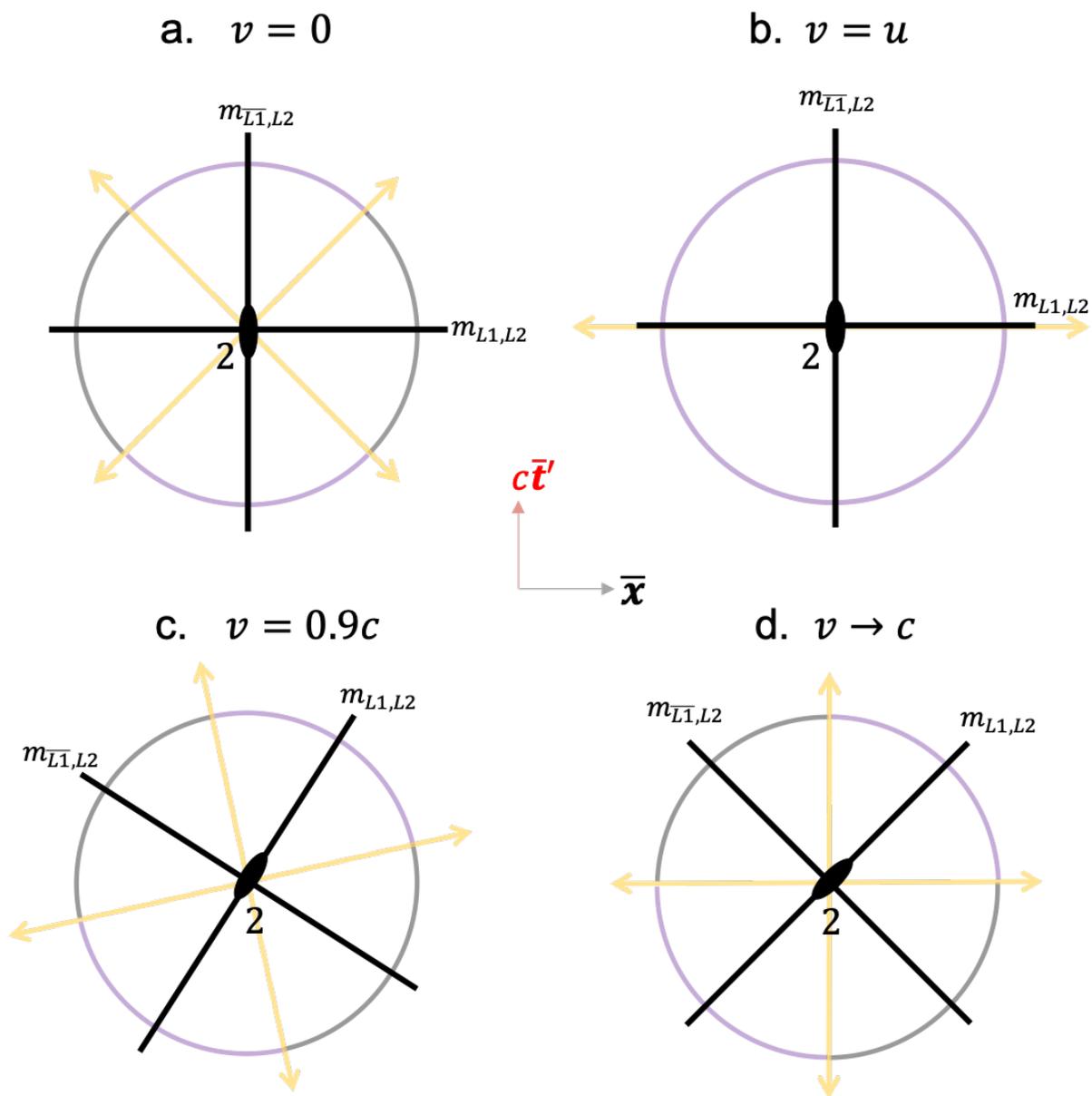

**Figure 11**: The RBS plots from **Figs. 6, 9b, 7, 8b** reproduced here in a lighter hue as panels a., b., c., and d., respectively. The symmetry elements of the extended RBS point group ***mm2*** are overlaid on each diagram indicating the 2-fold rotation at the center (black oval), and the two mirrors (black lines).

Four other subgroups of this symmetry group are also valid groups describing the 2D RBS, namely, point groups $m \equiv \{1, m_{L_1 L_2}\}$ or $\{1, m_{\overline{L_1} L_2}\}$, $2 \equiv \{1, 2\}$, and trivially $1 \equiv \{1\}$. Thus overall, there exist five 2D RBS colorless point groups: **mm2, m** (2 possibilities)**, 2** and **1**. Note that for **Figure 6** and **9b**, one of the mirrors is equivalent to the RBS space inversion in 2D (previously labeled $\overline{1}_{RBS}$). Similarly, the other mirror is equivalent to the RBS time reversal mentioned earlier ($1'_{RBS}$). Finally, the 2-fold is equivalent to the RBS spacetime reversal, $\overline{1}'_{RBS}$, mentioned earlier. Hence, one could alternately represent the **mm2** group for the cases of **Figure 6** and **9b** as $\mathbf{mm2} \equiv \{1, \overline{1}'_{RBS}, 1'_{RBS}, \overline{1}_{RBS}\}$. These groups therefore represent *extended RBS Lorentz groups*, $\mathcal{L}_{eRBS}$.

*Color symmetry including all the features of the RBS diagrams:* An antisymmetry such as time reversal, $1'$ will switch between two time-states, $t \leftrightarrow -t$ [18]. These two states can be associated with two colors, say black and purple, and thus $1'$ switches between black and purple colors representing the two-time states. Similarly, note that time-like and space-like events are distinguished by the parameter $\xi^2$ that switches sign from negative (time-like) to positive (space-like). If we introduce a new antisymmetry operation, $1^\xi$:

$$1^\xi : \xi^2 \leftrightarrow -\xi^2 \tag{18}$$

This operation thus switches the "color" between time-like (purple) and space-like (black) events. One could consider implementing this operation as follows: $1^\xi : x^2 \leftrightarrow -x^2$ and $1^\xi : t^2 \leftrightarrow -t^2$. An alternate way to perform this operation is $1^\xi : x \leftrightarrow t$. In either case, note that neither of these operations are elements of the $\mathcal{L}_{eRBS}$. Also note that $1^\xi$ is a self-inverse (i.e. $1^\xi \cdot 1^\xi = 1$), commutes with all the elements of the $\mathcal{L}_{eRBS}$ point groups mentioned earlier for the case of colorless symmetry groups that includes all the features of the RBS diagrams, and is not already an element of those groups. These are requirements for an operation to be an *antisymmetry* with respect to a group [18].

By performing the direct product $\mathcal{L}_{eRBS} \otimes \{1, 1^\xi\}$, one can generate *grey RBS symmetry groups* that explicitly contain $1^\xi$. (The "grey" is supposed to reflect a mixture of black and white (here purple is chosen instead of white) because of the explicit presence of $1^\xi$ that switches between the two colors). It's subgroups which do not explicitly contain $1^\xi$ are then called the *two-color RBS symmetry groups*. From **Figs. 6, 9b, 7, 8b**, we can conclude that $1^\xi$ is not explicitly present, i.e. swapping time- and space-like events will change the diagrams, hence it is not a symmetry element of the group. Hence grey RBS groups are excluded.

Next, we consider *two color RBS groups* in analogy with 2-color magnetic point groups [19]. **Figures 6, 9b, 7, 8b** exhibit the symmetry group $\mathbf{4^\xi mm^\xi}$. This is shown in **Figure 12**. The group elements are $\mathbf{4^\xi mm^\xi} \equiv \{1, 4^\xi, 4^{\xi^{-1}}, 2, m_{L1L2}, m_{\overline{L1}L2}, m_{L1}^\xi, m_{L2}^\xi\}$. The elements $4^\xi \equiv 4 \cdot 1^\xi$ and $m^\xi \equiv m \cdot 1^\xi$. The $4^\xi$ and $4^{\xi^{-1}}$ represent Euclidean 4-fold rotations by an angle of $\pm 2\pi/4$, respectively, followed by $1^\xi$. The colored mirrors $m_{L1}^\xi$ and $m_{L2}^\xi$ in **Figs. 6, 7** and **8b** are collinear with the two light lines in each figure. The uncolored mirrors, $m_{L1L2}$ and $m_{\overline{L1}L2}$ bisect the quadrants formed by the light lines. The subgroups of $\mathbf{4^\xi mm^\xi}$ such as $\mathbf{4^\xi} \equiv \{1, 4^\xi, 4^{\xi^{-1}}, 2\}$, $\mathbf{m} \equiv \{1, m_{L1L2}\}$ or $\{1, m_{\overline{L1}L2}\}$, and $\mathbf{m^\xi} \equiv \{1, m_{L1}^\xi\}$ or $\{1, m_{L2}^\xi\}$ are also allowed symmetry groups for this case. In the case of **Figure 9b**, there are no colored symmetry elements since all events are along time-like directions.

*Colorless symmetry ignoring the RBS light lines and the distinction between time- versus space-like events:* In such a case, the symmetry group is a Curie groups $\infty \mathbf{m} \equiv \{\mathbf{1}, \infty, m ...\}$ and its subgroup $\infty \equiv \{1, \infty, ..\}$ in 2D. The element $\infty$ represents an infinitesimal Euclidean rotation angle of $2\pi/\infty$ in the $(x, ct')$ plane. The element $m$ represents a vertical mirror in the plane. There are infinitely many such rotation and mirror elements in these groups, hence the "..." in the group.

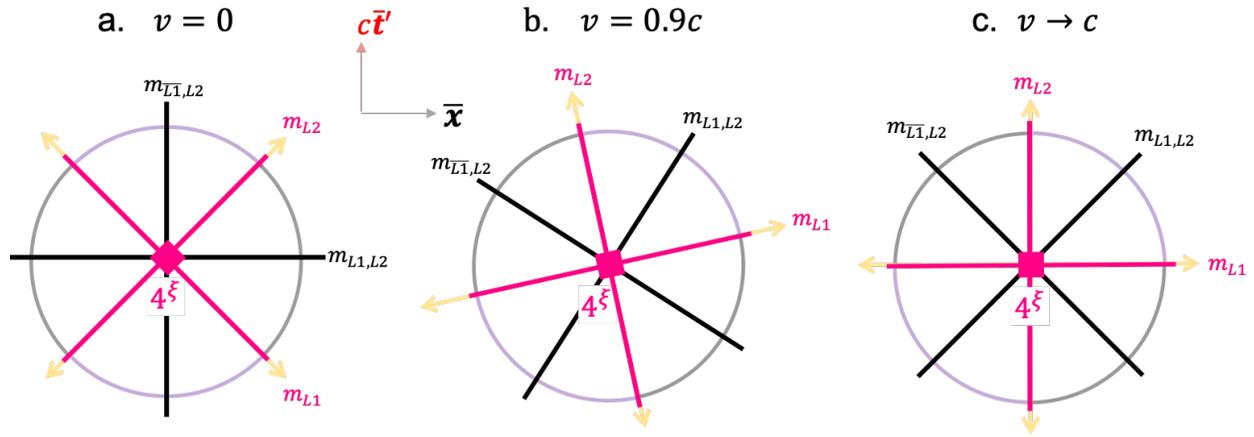

**Figure 12**: The 2D RBS plots from **Figs 6, 7, 8b** reproduced here in a lighter hue as panels a., b., and c., respectively. The symmetry elements of the extended 2-colored RBS point group $\mathbf{4^\xi mm^\xi}$ are overlaid on each diagram indicating the $4^\xi$ rotation axis at the center (red diamond), the two colorless mirrors (black lines), and the two-colored mirrors (red lines).

## 9. Three- and Four-Dimensional RBS point groups

Three dimensional RBS would have the coordinates of $(c\bar{t}', \bar{x}^1, \bar{x}^2)$, while 4D RBS would have the coordinates of $(c\bar{t}', \bar{x}^1, \bar{x}^2, \bar{x}^3)$. **Figure 13** depicts 3D RBS for two cases for (a) $v = 0$ and (b) $v = u$, similar to the 2D RBS in **Figures 6** and **9b**, respectively. The Curie group is $\frac{\infty}{m}m$ for both cases. In both cases, there is one $\infty$-fold axis and horizontal mirror (*m* in the denominator) in the equatorial plane perpendicular to the $\infty$-fold axis as shown in **Figure 13c**. There are infinitely many vertical mirrors (*m* in the numerator); one of them is depicted in panel c, and an infinite number of vertical mirrors are generated by the $\infty$-fold axis. One 2-fold rotation axis is depicted and again there are infinitely many 2-folds generated by the $\infty$-fold axis. A series of events in the form of a blue ring (a flock of birds forming a ring?) in the upper and lower hemispheres is shown in panel **Figure 13c** reflecting the $\frac{\infty}{m}m$ symmetry.

Existing symmetries of the isotropic 3D RBS can be broken by arranging various events in the 3D RBS so as to break certain symmetries and create RBS crystals with lower symmetry. The following Curie subgroups of $\frac{\infty}{m}m$ are also valid groups describing 3D RBS if some symmetries are broken: $\frac{\infty}{m}$, $\infty m$, $\infty 2$, and $\infty$ [20]. For example, by placing a single event in the upper hemisphere in **Fig. 13 a** or **b** and nowhere else would break all the symmetries depicted in **Fig. 13c**; it would correspond to the 3D point group labeled **1** whose only element is identity, 1. By placing two events, one related to the other by 3D RBS inversion, $\bar{1}'_{RBS}: (c\bar{t}', \bar{x}^1, \bar{x}^2) \rightarrow -(c\bar{t}', \bar{x}^1, \bar{x}^2)$, one obtains the 3D RBS group $\bar{\mathbf{1}}'_{RBS} \equiv \{1, \bar{1}'_{RBS}\}$ as shown in **Figure 14a** for the $v = u$ case from **Figure 13b**.

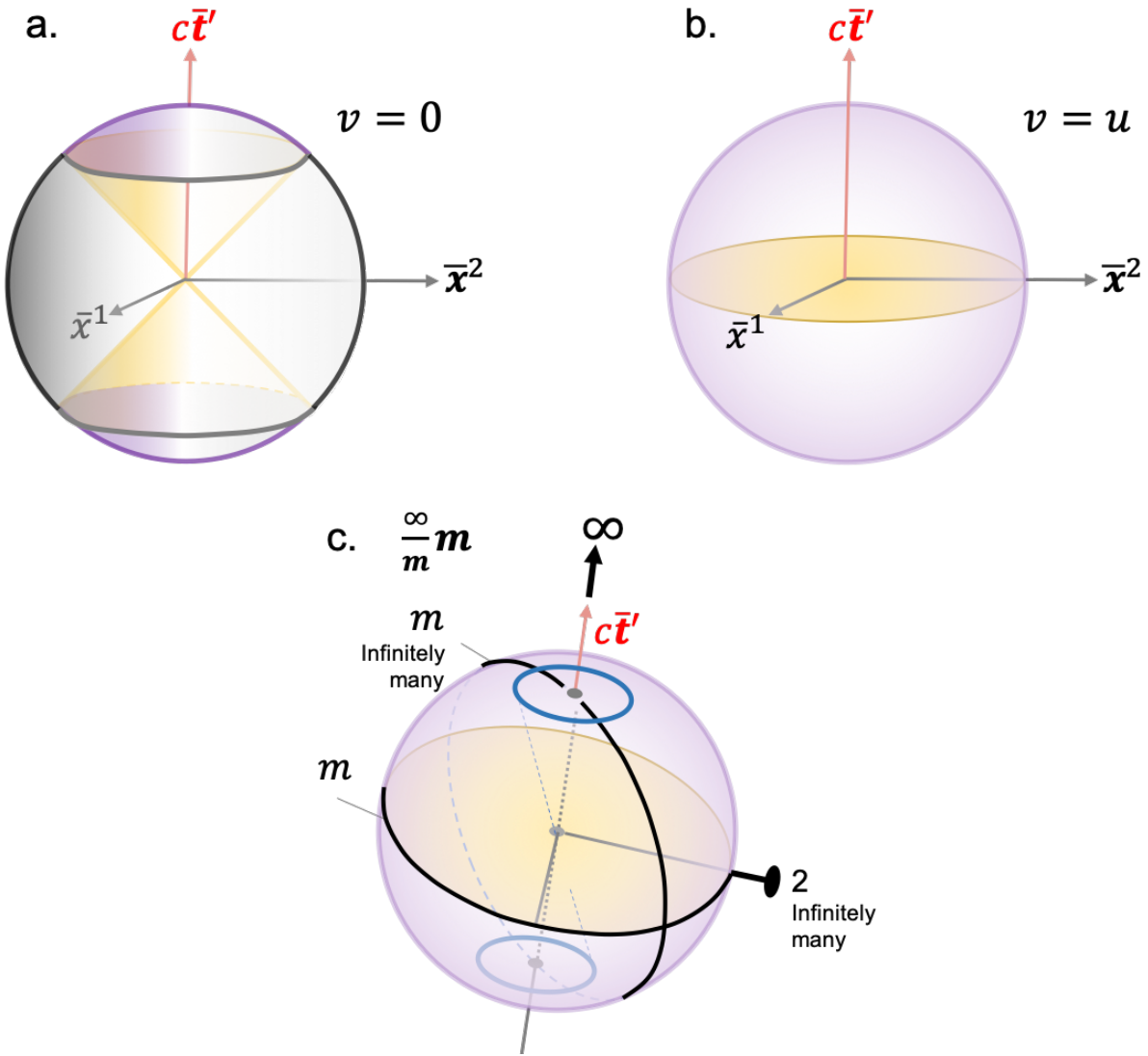

**Figure 13**: Isotropic 3D RBS coordinates depicted for (a) $v = 0$ and (b) $v = u$, similar to the 2D RBS in **Figures 6** and **9b**, respectively. The grey versus purple sphere surfaces indicate space-like versus time-like events, respectively. The yellow light cones are depicted in (a), while the light plane is depicted in (b) as the equatorial plane. Panel (c) depicts one ∞-fold rotation axis, one (of infinitely many) 2-fold rotation axis, one horizontal mirror and one (of infinitely many) vertical mirror. The 3D Curie point group for both (a) and (b) is $\frac{\infty}{m}m$.

The ∞-fold can be replaced by a *p*- fold rotation (*p* is a natural number) using appropriately placed events. If one restricts themselves to *periodic* 3D space crystals, only 1, 2, 3, 4 or 6-fold rotation axes are allowed [20]. **Figure 14** shows events placed as blue ovals on the surface of an RBS surface for the $v = u$ case (shown in **Figure 13b**) in order to create six of the seven *holohedral* point groups in periodic 3D space crystals now applied to 3D RBS: $\bar{1}'_{RBS}$, **2/m, mmm, 4/mmm**, $\bar{3}m$, and **6/mmm**. (The only missing holohedral group in **Figure 14** is the cubic group ***m3m*** which is not consistent with the 3D RBS. This is because in breaking symmetry through the placement of events, some aspects of the RBS are "baked in" and cannot be changed, such as the RBS light lines, planes and cones, and the resulting "crease" between the time-like and space-like events as seen in **Figure 13a** and **b**.). All other RBS point groups are subgroups of these six RBS holohedral groups [20]. None of the 14 conventional colored 3D Curie groups listed in Ref. [21] can be associated with the 3D RBS structures in **Figure 14** by the inclusion of $1^\xi$. Since $1^\xi$ results in "dissolving" and "reforming" the light cones, and the crease between time-like and space-like event surfaces in **Figure 13a**, it does not conform to the definition of a typical symmetry element where no cuts or stitches to the object in question are allowed; that is the domain of topological distortions, and hence are not discussed further here.

One can construct similar 4D RBS structures and the corresponding Curie groups. All the point groups and space groups for space crystals in 4D are listed in literature [15]. The group $\frac{\infty}{m}m$ in 4D would be valid, except *m* would represent a *hyperplane* (of dimension 3) in 4D. For the case of $v = u$ for the 4D $(c\bar{t}', \bar{x}^1, \bar{x}^2, \bar{x}^3)$ coordinates, the horizontal 4D hyperplane mirror perpendicular to the ∞-fold rotation axis will be given by the diagonal tensor [-1, 1, 1, 1] (which is equivalent to the RBS time reversal in 4D). One of the vertical 4D hyperplane mirrors parallel to the ∞-fold would be, for example, the diagonal matrix given by [1, -1, 1, 1] perpendicular to the $\bar{x}^1$ axis. The ∞-fold axis parallel to $\bar{t}'$ axis would rotate the stated vertical hyperplane mirror to generate infinitely many of them. The subgroups of this group would again be valid descriptions of the RBS. Crystallographic 4D RBS groups can also be deduced from the well enumerated 4D space crystallographic groups listed in literature [15].

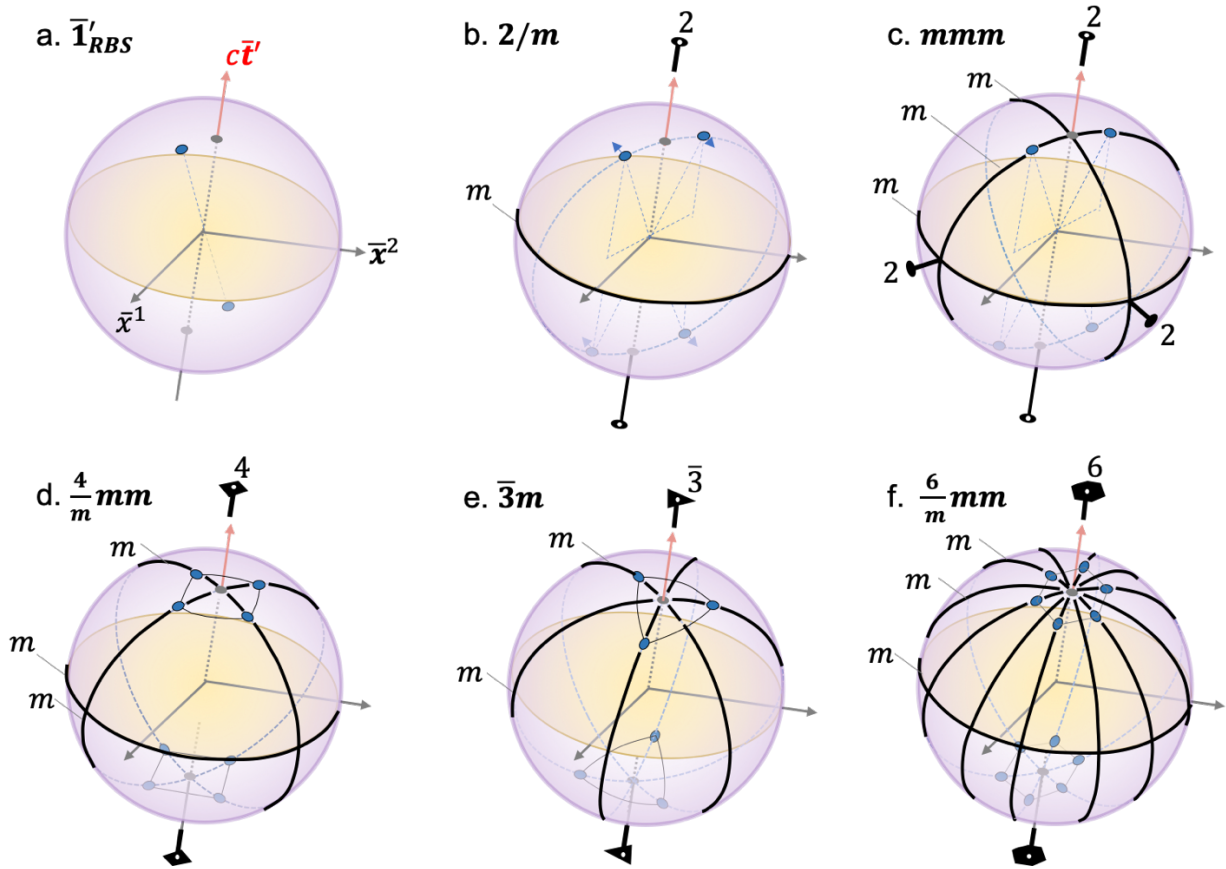

**Figure 14**: Six 3D holohedral RBS point groups for periodic RBS crystals. The sphere from **Figure 13c** for the case of $v = u$ is shown in each panel, with appropriately placed events (blue ovals) on each hemisphere to break specific symmetries and retain others. The blue arrows associated with the events in panel b suggests a series of additional events stretching in the direction of the arrows. The generating symmetry elements for each group are indicated.

## 10. Periodic RBS crystals

The defining feature periodic spatial crystals is their translational symmetry, namely, that they are periodic in various spatial dimensions. In describing their symmetry, one moves beyond point groups to add translations to create space groups [22,23]. In the context of conventional Minkowski spacetime, one moves from Lorentz groups to Poincare' groups. The group theoretical procedure to move from point groups to space groups is well established [22]. Here, given the equivalence established between space crystals and RBS crystals, one could similarly move from the RBS point groups to RBS Poincare' groups in analogy with space groups. Below, we limit our discussion to 2D, but similar extensions will be possible in higher dimensions.

There are 17 2D space group types describing spatial crystals [24]. In order to keep the discussion simple, let us focus on the simplest case of $v = u$ depicted by the 2D RBS group depicted in **Figure 9b**, where the blended coordinates are between the GF and the event BF. Since the light line is parallel to the space axis, $\bar{x}$ in this case, and the resulting symmetry as seen before is ***mm*2**, let us restrict our discussion to space groups whose site symmetries (point group symmetries at individual locations within the crystal) are restricted to point group symmetries of ***mm*2** or its subgroups. **Figure 15** shows such 2D RBS space groups, where the group labels are picked to be synonymous with the corresponding 2D space group labels for space crystals.

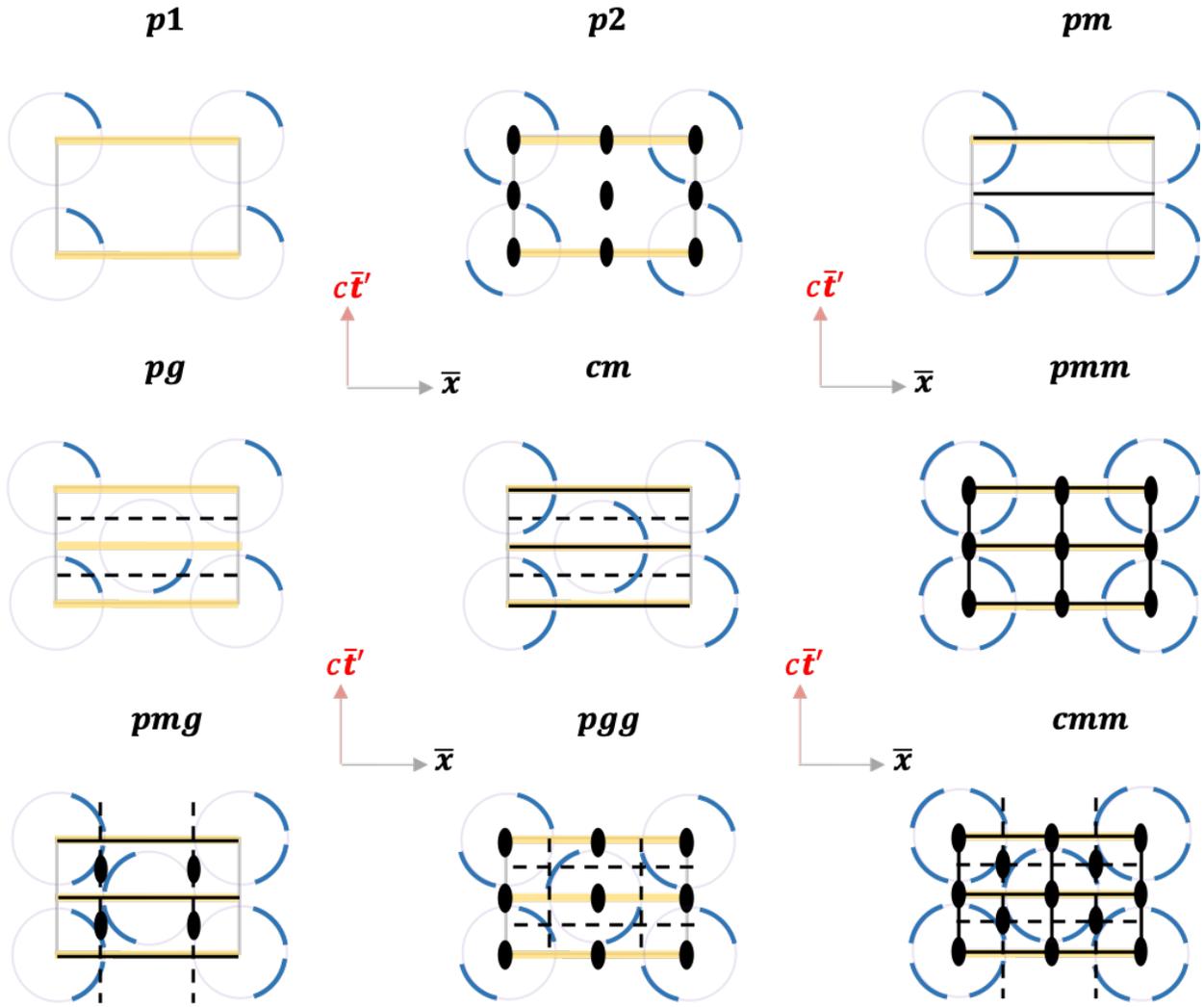

**Figure 15**: Examples of 2D RBS space groups adopting the same labels as the corresponding 2D space groups for space crystals. A unit cell is shown in each case by a grey rectangle. The faded purple circle at each lattice site is the same as the RBS circle in **Figure 9b** for the case of $v = u$. The blue arcs represent a series of events (an RBS spacetime flock of birds?) being observed in RBS coordinates as indicated by the axes $(\bar{x}, \bar{t}')$. Dashed black lines are glide planes, solid black lines are mirrors, yellow lines are light lines, and the black ovals represent 2-fold rotation axes.

The RBS crystals can be imagined as a series of events periodically arranged in the RBS being observed by an RBS observer at the origin. In the 2D case, the periodicity arises from translations along the $\bar{x}$ and the $\bar{t}'$ axes. Naturally, the event periodicity will result in the RBS observer herself being replicated periodically in the RBS as depicted. Glide planes (dashed lines in **Fig. 15**) can be observed now where one mirrors across the glide plane, and then translate by half a unit cell along the glide plane. These types of symmetries are not obvious in the conventional MS constructions of spacetime depicted in Fig. 2.

How about 2D space groups with say 3- 4- and 6-fold rotations? These are excluded in the case of a fixed relative orientation of the light lines in the RBS; higher fold rotations than 2-fold will rotate the RBS light lines as well, and hence these RBS space groups will have to be composed with varying $v$ in the RBS. Similar constructions can be made in 3D and 4D RBS. These are interesting topics left to be explored in future works.

## 11. Conclusion

In conclusion, while time crystals are of great current interest [25,26], this work extends the concept to relativistic spacetime crystals. By considering blended inertial frames between two inertial observers and then renormalizing the coordinates of an event observed by them by $\chi$ (which is a function, given in E. (8), of the relative velocities between the ground and the train frames, $v$, and between the ground and the event frames, $u$), one can generate the RBS coordinates $(c\bar{t}', \bar{x})$ and $(c\bar{t}, \bar{x}')$. These coordinates transform the hyperbolic geometry of the Minkowski (MS) spacetime into a renormalized Blended Spacetime (RBS) that exhibits a Euclidean construction. The Lorentz boosts become continuous Euclidean rotations, and the RBS geometry also exhibits a new set of light lines. Mapping between the MS and the RBS frames shows that they have equivalent relativistic physics content: every point not on the light lines in MS maps to a unique point in RBS. Every point on the light lines in MS maps to the origin in RBS. Conversely, the light lines in the RBS map to the $+(\infty, \pm\infty)$ or $-(\infty, \pm\infty)$ limits of the light lines in the MS. Points not on the light lines in the RBS uniquely map to points in the MS.

These mappings between MS and RBS give rise to equivalent representations of the relativistic physics in both descriptions. This is based on based on three considerations: (1) the equivalence mapping in Figure 10 between the MS and RBS coordinates. (2) Einstein's first and second postulates hold still. Blending of the frames does not modify them, since one can always revert back from the RBS to the MS coordinates and recover these postulates. (3) Lorentz transformation (Eq. (3)) and the invariance of the spacetime interval, $\xi^2$ (Eq. (2)) are still valid, since the equivalent RBS statements in Eq. (11c) and (12), respectively, were derived from them.

However, mathematically speaking, the Euclidean geometry in RBS allows one to smoothly mathematically "cross" the RBS light lines, which is not possible in the hyperbolic geometry in MS. This feature allows us to write Lorentz boosts as Euclidean rotations, which in turn helps map the Lorentz group of the RBS to equivalent crystallographic symmetry groups already well known in space crystals. The RBS point groups in 2D, 3D and 4D are identified to be those associated with cylinders in various dimensions: rectangle in 2D, cylinder in 3D, and hypercylinder in 4D. With the addition of translations, examples are given for 2D RBS space groups that describe RBS crystals; RBS space groups of higher dimensions should be straightforward in a similar manner. A *Mathematica* file is provided for a reader to plot the MS and RBS constructions for themselves.

On a more general mathematical note, this approach could allow one to straddle between Euclidean and hyperbolic coordinate systems in a flat space or spacetime. For a set of $n$ linearly independent coordinates $x^\mu$, $\mu: 1-n$, if the Eigen value of the metric tensor for the first $k$ coordinates is -1, and that for the remaining ($n$-$k$) coordinates is +1, and if a linear transformation between $x^\mu$ and $(x^\mu)'$ coordinates exists that leaves the interval $(x^1)^2 + (x^1)^2 + \cdots + (x^k)^2 - (x^{k+1})^2 - (x^{k+2})^2 - (x^n)^2$ invariant before and after the transformation, then one can define a blended coordinate system between primed and unprimed coordinates with a Euclidean interval $(x^1)^2 + (x^1)^2 + \cdots + (x^k)^2 + (x^{k+1})'^2 + (x^{k+2})'^2 + \cdots + (x^n)'^2 = \zeta^2$. If $\bar{x}^\mu = x^\mu/\zeta$, is defined, then $\sum_\mu \bar{x}^{\mu 2} = 1$ is a unit circle in a Euclidean frame. Going forward, it will be interesting to explore quasi-1D RBS magnetic groups, periodic and aperiodic RBS crystallographic groups in various dimensions, RBS quasicrystals, and the full scope of the renormalized blended frames in covariant electrodynamics, relativistic physics and quantum gravity [27]. **Appendix B** provides a preliminary sketch for how one might consider extensions of this work to general relativity.

**APPENDIX A**

Below are *Mathematica®* codes used to generate select figures. Copy and paste them in a *Mathematica®* notebook and then execute them. A supplementary *Mathematica®* notebook file and its PDF export are also provided with the same content.

**Figure 2**

```
Clear[\[Beta], \[Xi]]
ParametricPlot[{{Sinh[\[Beta]],
  Cosh[\[Beta]]}, {-Sinh[\[Beta]], -Cosh[\[Beta]]}, {Cosh[\[Beta]],
  Sinh[\[Beta]]}, {-Cosh[\[Beta]], -Sinh[\[Beta]]}, {\[Xi], \[Xi]}, \
{\[Xi], -\[Xi]}, {\[Xi]*Sinh[ArcTanh[0.9]], \[Xi]*
  Cosh[ArcTanh[0.9]]}, {\[Xi]*Cosh[ArcTanh[0.9]], \[Xi]*
  Sinh[ArcTanh[0.9]]}}, {\[Beta], -ArcTanh[0.999],
```

ArcTanh[0.999]}, {\[Xi], -3, 3}, PlotRange -> {{-3, 3}, {-3, 3}}]

**Figure 3**

Clear[\[Theta], \[Phi], \[Xi]]

\[Theta] = ArcSin[0]

p1 = ParametricPlot[{{Sin[\[Phi]]/Cos[\[Phi] - \[Theta]], Sqrt[

  Abs[Sec[\[Theta]]*Sec[2*\[Phi] - \[Theta]]]]} }, {\[Phi],

  ArcSin[-0.99999], ArcSin[0.99999]}, PlotRange -> {{-1, 1}, {0, 10}}]

\[Theta] = ArcSin[0.9]

p2 = ParametricPlot[{{Sin[\[Phi]]/Cos[\[Phi] - \[Theta]], Sqrt[

  Abs[Sec[\[Theta]]*Sec[2*\[Phi] - \[Theta]]]]} }, {\[Phi],

  ArcSin[-0.99999], ArcSin[0.99999]}, PlotRange -> {{-1, 1}, {0, 10}}]

\[Theta] = \[Phi]

p3 = ParametricPlot[{{Sin[\[Phi]]/Cos[\[Phi] - \[Theta]], Sqrt[

  Abs[Sec[\[Theta]]*Sec[2*\[Phi] - \[Theta]]]]} }, {\[Phi],

  ArcSin[-0.99999], ArcSin[0.99999]}, PlotRange -> {{-1, 1}, {0, 10}}]

Show[p1, p2, p3, PlotTheme -> "Minimal" ]

**Figure 4**

Clear[\[Theta], \[Phi], \[Xi]]

\[Theta] = ArcSin[0]

\[Xi] = 1

\[Chi] = Sqrt[Abs[Sec[\[Theta]]*Sec[2*\[Phi] - \[Theta]]]]

x = \[Xi]*\[Chi]*Sin[\[Phi]]

tp = \[Xi]*\[Chi]*Cos[\[Phi]]

ParametricPlot[{{x,

  tp}, {\[Xi], \[Xi]}, {\[Xi], -\[Xi]}}, {\[Phi], -Pi,

  Pi}, {\[Xi], -5, 5}, PlotPoints -> 50,

PlotRange -> {{-3, 3}, {-3, 3}}]

## Figure 5

Clear[\[Theta], \[Phi], \[Xi]]

\[Theta] = ArcSin[0.9]

\[Xi] = 1

\[Chi] = Sqrt[Abs[Sec[\[Theta]]*Sec[2*\[Phi] - \[Theta]]]]

x = \[Xi]*\[Chi]*Sin[\[Phi]]

tp = \[Xi]*\[Chi]*Cos[\[Phi]]

ParametricPlot[{{x,

  tp}, {\[Xi], \[Xi]*0.229416}, {\[Xi], -\[Xi]*4.358898943540674`}}, \
 
{\[Phi], -Pi, Pi}, {\[Xi], -5, 5}, PlotRange -> {{-5, 5}, {-5, 5}}]

## Figure 6

Clear[\[Theta], \[Phi], \[Xi], \[Chi], x, tp]

\[Theta] = ArcSin[0]

\[Xi] = 1

\[Chi] = Sqrt[Abs[Sec[\[Theta]]*Sec[2*\[Phi] - \[Theta]]]]

xn = \[Chi]*Sin[\[Phi]]/\[Chi]

tpn = \[Chi]*Cos[\[Phi]]/\[Chi]

ParametricPlot[{{xn,

  tpn}, {\[Xi], \[Xi]}, {\[Xi], -\[Xi]}}, {\[Phi], -Pi,

 Pi}, {\[Xi], -2, 2}, PlotRange -> {{-2, 2}, {-2, 2}}]

## Figure 7

Clear[\[Theta], \[Phi], \[Xi], \[Chi], x, tp]

\[Theta] = ArcSin[0.9]

\[Xi] = 1

\[Chi] = Sqrt[Abs[Sec[\[Theta]]*Sec[2*\[Phi] - \[Theta]]]]

xn = \[Chi]*Sin[\[Phi]]/\[Chi]

tpn = \[Chi]*Cos[\[Phi]]/\[Chi]

ParametricPlot[{{xn,

  tpn}, {\[Xi], \[Xi]*0.229416}, {\[Xi], \

-\[Xi]*4.358898943540674`}}, {\[Phi], -Pi, Pi}, {\[Xi], -2, 2},

 PlotRange -> {{-2, 2}, {-2, 2}}]

**Figure 8a**

Clear[\[Theta], \[Phi], \[Xi]]

\[Theta] = ArcSin[0.99999]

\[Xi] = 1

\[Chi] = Sqrt[Abs[Sec[\[Theta]]*Sec[2*\[Phi] - \[Theta]]]]

x = \[Xi]*\[Chi]*Sin[\[Phi]]

tp = \[Xi]*\[Chi]*Cos[\[Phi]]

ParametricPlot[{{x, tp}}, {\[Phi], -Pi, Pi},

 PlotRange -> {{-40, 40}, {-40, 40}}]

**Figure 8b**

Clear[\[Theta], \[Phi], \[Xi], \[Chi], x, tp]

\[Theta] = ArcSin[0.99999]

\[Xi] = 1

\[Chi] = Sqrt[Abs[Sec[\[Theta]]*Sec[2*\[Phi] - \[Theta]]]]

xn = \[Chi]*Sin[\[Phi]]/\[Chi]

tpn = \[Chi]*Cos[\[Phi]]/\[Chi]

ParametricPlot[{{xn, tpn}}, {\[Phi], -Pi, Pi},

 PlotRange -> {{-2, 2}, {-2, 2}}]

**Figure 9a**

```
Clear[\[Theta], \[Phi], \[Xi]]

\[Theta] = \[Phi]

\[Xi] = 1

\[Chi] = Sqrt[Abs[Sec[\[Theta]]*Sec[2*\[Phi] - \[Theta]]]]

x = \[Xi]*\[Chi]*Sin[\[Phi]]

tp = \[Xi]*\[Chi]*Cos[\[Phi]]

ParametricPlot[{{x, tp}}, {\[Phi], -Pi, Pi},

 PlotRange -> {{-5, 5}, {-1.5, 1.5}}]
```

**Figure 9b**

```
Clear[\[Theta], \[Phi], \[Xi], \[Chi], x, tp]

\[Theta] = \[Phi]

\[Xi] = 1

\[Chi] = Sqrt[Abs[Sec[\[Theta]]*Sec[2*\[Phi] - \[Theta]]]]

xn = \[Chi]*Sin[\[Phi]]/\[Chi]

tpn = \[Chi]*Cos[\[Phi]]/\[Chi]

ParametricPlot[{{xn, tpn}}, {\[Phi], -Pi, Pi},

 PlotRange -> {{-2, 2}, {-2, 2}}]
```

**APPENDIX B**: **Sketch of Blended coordinates in the Rindler and Schwarschild geometries**

The line element in Rindler Geometry in a flat 2D spacetime is given by the *differential* line element, $ds^2 = dx^2 - c^2 dt^2 = d\xi^2 - \xi^2 d\beta^2$ which captures many of the same properties as the Schwarzschild geometry in General Relativity. [28] The second equality uses the hyperbolic polar coordinates $(\xi, \beta)$ shown in **Figure 2**, which is also called the Rindler coordinates. Upon computing $ds'^2 = dx'^2 - c^2 dt'^2$ in the train inertial frame (TF) moving at a *constant* relative speed of $v = c\tanh\alpha$ with respect to the ground frame, one can show that $ds'^2 = d\xi^2 - \xi^2 d(\beta - \alpha)^2 = d\xi^2 - \xi^2 d\beta^2$, since $\alpha$ is assumed constant; thus, $ds^2 = ds'^2$ is an invariant.

If we now define a Rindler differential line element in the blended frame as $ds^{o\prime 2} = dx^2 + c^2 dt'^2 = dx'^2 + c^2 dt^2 = ds'^{o2}$, then one can show that $ds^{o\prime 2} = \chi^2(d\xi^2 + \xi^2 d\beta^2) + \kappa^2 \xi d\xi d\beta$, where $\kappa^2 = \sinh 2\beta + \sinh(2\beta - 2\alpha)$. Both the factors $\chi$ and $\kappa$ are functions of $\alpha$ that determines the relative speeds of the two inertial frames (GF and TF).

Consider two special cases in the Rindler geometry above: a constant acceleration ($d\xi = 0$), and (trivially) a constant velocity ($d\beta = 0$) of the bird. In the former case ($d\xi = 0$), $ds^{o\prime 2} = \chi^2 \xi^2 d\beta^2$, and thus one could define renormalized blended coordinates, $d\bar{x} = (1/\chi\xi)(dx/d\beta)$ and $cd\bar{t}' = (c/\chi\xi)(dt'/d\beta)$, such that $d\bar{x}^2 + c^2 d\bar{t}'^2 = 1$, a unit circle for any worldline in the Rindler geometry with a constant acceleration. In the latter case ($d\beta = 0$), $ds^{o\prime 2} = \chi^2 d\xi^2$, and one could define renormalized blended coordinates, $d\bar{x} = (1/\chi)(dx/d\xi)$, and $cd\bar{t}' = (c/\chi)(dt'/d\xi)$, to again recover a unit circle. More generally, one could define $d\bar{x} = dx/ds^{o\prime}$, and $cd\bar{t}' = dt'/ds^{o\prime}$, such that $d\bar{x}^2 + c^2 d\bar{t}'^2 = 1$ for any worldline in the Rindler geometry.

Now let us consider the curved spacetime. Schwarzchild metric describes the gravitational field of a point mass, $m$, at the origin; it is a spherically symmetric solution of Einstein's equation in vacuum. [28] The line element is given in polar coordinates, $(r, \theta, \phi)$ with the origin centered at the mass by

$$ds^2 = -\sigma^{t2} + \sigma^{r2} + r^2 d\theta^2 + r^2 \sin^2\theta d\phi^2 \qquad (19)$$

Where $\sigma^{t^2} = (1 - 2m/r)dt^2$ and $\sigma^{r^2} = dr^2/(1 - 2m/r)$, where the abbreviation, $ct \to t$ and $mG/c^2 \to m$ has been used. As a specific example, consider a *shell observer* sitting on an imaginary shell at a radius $r$ from the mass, on the equator at a fixed $\theta = \pi/2$, $(d\theta = 0)$ and a fixed azimuth ($d\phi = 0$). Note that as $r \to \infty$, this metric reduces to that of the Minkowski metric of flat spacetime. Consider the rain coordinates of a radial geodesic (for example, a radial worldline from $r \to \infty$ towards $r \to 0$). The relative speed of the radial observer as she crosses the shell observer can be shown to be $v = ctanh\alpha = -c\sqrt{2m/r}$, where the minus sign indicates motion in the $-\hat{r}$, or the inward radial direction. A Lorentz transformation between the shell coordinates, $(\sigma^t, \sigma^r)$ and the rain coordinates, $(\sigma^T, \sigma^R)$ is given by $\begin{pmatrix}\sigma^T \\ \sigma^R\end{pmatrix} = \Lambda \begin{pmatrix}\sigma^t \\ \sigma^r\end{pmatrix}$, where

$$\sigma^T = dt + \frac{dr\sqrt{\frac{2m}{r}}}{\left(1-\frac{2m}{r}\right)}, \text{ and}$$

$$\sigma^R = \sqrt{\frac{2m}{r}}dt + \frac{dr}{\left(1-\frac{2m}{r}\right)} \quad (20)$$

Further, the metric is invariant, namely, $ds^2 = -\sigma^{t^2} + \sigma^{r^2} = -\sigma^{T^2} + \sigma^{R^2}$. If we now consider a blended reference frame between the shell and the rain coordinates, then, $ds^{o\prime 2} = \sigma^{T^2} + \sigma^{r^2} = \sigma^{t^2} + \sigma^{R^2} = ds^{\prime o^2}$. Rearranging, we can rewrite this as $(\sigma^T/ds^{o\prime})^2 + (\sigma^r/ds^{o\prime})^2 = 1$, where $\bar{\sigma}^T = \sigma^T/ds^{o\prime}$ and $\bar{\sigma}^r = \sigma^r/ds^{o\prime}$. Thus, in principle, blended renormalized Euclidean coordinates are possible locally on a manifold in General Relativity.

**Acknowledgments:**

VG would like to acknowledge support from the National Science Foundation grant number DMR-1807768. Discussions with Martin Bojowald, Matijn Van Kuppeveld, Haricharan Padmanabhan, Vincent S. Liu and Zhiwen Liu is gratefully acknowledged.